\documentclass[twocolumn]{aastex63}

\usepackage{multirow}
\usepackage{color}
\usepackage{amsmath,amsfonts,bm}
\usepackage{hyperref}
\usepackage[T1]{fontenc}
\usepackage{graphicx}	
\usepackage{dcolumn}
\usepackage{lineno}
\usepackage{seqsplit}
\usepackage{bm}
\usepackage{amsmath}

\newcommand{\planck}{{\it Planck}}
\newcommand{\snr}{$S/N$}
\newcommand{\smica}{\texttt{SMICA}}

\newcommand{\healpix}{\texttt{HEALPix}}
\newcommand{\takasims}{Takahashi simulations}

\newcommand{\howmanyvoidsintotal}{7378}

\newcommand{\boxsize}{\mbox{$10^{\circ} \times 10^{\circ}$}}
\newcommand{\thetabinsize}{0.25^{\circ}}

\newcommand{\howmanysigma}{5.3}

\newcommand{\lmaxforanalysis}{2048}
\newcommand{\lv}{\lambda_{v}}
\newcommand{\Al}{A_{L}}

\newcommand{\rvunits}{h^{-1}{\rm Mpc}}
\newcommand{\totalvoids}{7378}

\newcommand{\wffirstbinAl}{0.73 \pm 0.27}
\newcommand{\wfsecondbinAl}{1.37 \pm 0.54}
\newcommand{\wfthirdbinAl}{0.96 \pm 0.76}
\newcommand{\wffourthbinAl}{1.38 \pm 0.52}
\newcommand{\wffifthbinAl}{0.86 \pm 0.27}
\newcommand{\wfcombinedAl}{0.97 \pm 0.19}
\newcommand{\wfallvoidsAl}{1.33 \pm 0.35}
\newcommand{\wfcombinedsnr}{5.1}

\newcommand{\wfcombinedlargescalesfilteredAl}{1.01 \pm 0.19}
\newcommand{\wfcombinednaiveAl}{0.64 \pm 0.13}
\newcommand{\mfcombinedlargescalesfilteredAl}{1.17 \pm 0.21}

\newcommand{\wfallvoidsbiasnoerrorbar}{3.22}
\newcommand{\wfallinvvoidsbias}{0.32 \pm 0.06}

\newcommand{\noSZwffirstbinAl}{0.73 \pm 0.32}
\newcommand{\noSZwfsecondbinAl}{0.46 \pm 0.62}
\newcommand{\noSZwfthirdbinAl}{1.68 \pm 0.80}
\newcommand{\noSZwffourthbinAl}{1.47 \pm 0.61}
\newcommand{\noSZwffifthbinAl}{0.72 \pm 0.32}
\newcommand{\noSZwfcombinedAl}{0.80 \pm 0.22}
\newcommand{\noSZwfallvoidsAl}{1.03 \pm 0.42}

\newcommand{\noSZmffirstbinAl}{1.08 \pm 0.34}
\newcommand{\noSZmfsecondbinAl}{1.14 \pm 0.64}
\newcommand{\noSZmfthirdbinAl}{1.17 \pm 0.89}
\newcommand{\noSZmffourthbinAl}{1.30 \pm 0.64}
\newcommand{\noSZmffifthbinAl}{1.23 \pm 0.34}
\newcommand{\noSZmfcombinedAl}{1.16 \pm 0.34}
\newcommand{\noSZmfallvoidsAl}{0.97 \pm 0.44}

\newcommand{\nullwffirstbinAl}{-0.33 \pm 0.27}
\newcommand{\nullwfsecondbinAl}{-0.25 \pm 0.52}
\newcommand{\nullwfthirdbinAl}{-0.03 \pm 0.74}
\newcommand{\nullwffourthbinAl}{-0.28 \pm 0.49}
\newcommand{\nullwffifthbinAl}{0.31 \pm 0.26}
\newcommand{\nullwfcombinedAl}{0.07 \pm 0.19}
\newcommand{\nullwfallvoidsAl}{0.59 \pm 0.37}

\newcommand{\nfcombinedAl}{0.98 \pm 0.20}

\newcommand{\mffirstbinAl}{0.95 \pm 0.28}
\newcommand{\mfsecondbinAl}{1.46 \pm 0.53}
\newcommand{\mfthirdbinAl}{0.60 \pm 0.76}
\newcommand{\mffourthbinAl}{1.27 \pm 0.52}
\newcommand{\mffifthbinAl}{1.20 \pm 0.28}
\newcommand{\mfcombinedAl}{1.10 \pm 0.21}
\newcommand{\mfallvoidsAl}{1.16 \pm 0.37}
\newcommand{\mfcombinedsnr}{5.3}

\defcitealias{takahashi17}{T17}

\newcommand{\abstracttext}
{
We report a $\howmanysigma\sigma$ detection of the gravitational lensing effect of cosmic voids from the Baryon Oscillation Spectroscopic (BOSS) Data Release 12 seen in the \planck{} 2018 cosmic microwave background (CMB) lensing convergence map. To make this detection, we introduce new optimal techniques for void stacking and filtering of the CMB maps, such as binning voids by a combination of their observed galaxy density and size to separate those with distinctive lensing signatures. We calibrate theoretical expectations for the void-lensing signal using mock catalogs generated in a suite of 108 full-sky lensing simulations from \citet{takahashi17}. Relative to these templates, we measure the lensing amplitude parameter in the data to be $A_L=1.10\pm0.21$ using a matched-filter stacking technique, and confirm it using an alternative Wiener filtering method. We demonstrate that the result is robust against thermal Sunyaev-Zel'dovich contamination and other sources of systematics. We use the lensing measurements to test the relationship between the matter and galaxy distributions within voids, and show that the assumption of linear bias with a value consistent with galaxy clustering results is discrepant with observation at $\sim 3\sigma$; we explain why such a result is consistent with simulations and previous results, and is expected as a consequence of void selection effects. We forecast the potential for void-CMB lensing measurements in future data from the Advanced ACT, Simons Observatory and CMB-S4 experiments, showing that, for the same number of voids, the achievable precision improves by a factor of more than two compared to \planck{}.
}

\begin{document}

\title{The Gravitational Lensing Signatures of BOSS Voids in the Cosmic Microwave Background}
\shorttitle{CMB Lensing of BOSS Voids}
\shortauthors{Raghunathan, Nadathur, Sherwin \& Whitehorn}

\author[0000-0003-1405-378X]{Srinivasan Raghunathan} 
\affiliation{Department of Physics and Astronomy, University of California, Los Angeles, CA 90095, USA}

\author[0000-0001-9070-3102]{Seshadri Nadathur} 
\affiliation{Institute of Cosmology and Gravitation, University of Portsmouth, Burnaby Road, Portsmouth PO1 3FX, United Kingdom}

\author[0000-0002-4495-1356]{Blake D. Sherwin} 
\affiliation{Kavli Institute for Cosmology, University of Cambridge, Madingley Road, Cambridge CB3 OHA, United Kingdom}
\affiliation{Department of Applied Mathematics and Theoretical Physics, University of Cambridge, Wilberforce Road, Cambridge CB3 0WA, UK}

\author[0000-0002-3157-0407]{Nathan Whitehorn} 
\affiliation{Department of Physics and Astronomy, University of California, Los Angeles, CA 90095, USA}


\correspondingauthor{Srinivasan Raghunathan}\email{sri@physics.ucla.edu}
\correspondingauthor{Seshadri Nadathur}\email{seshadri.nadathur@port.ac.uk}




\begin{abstract}
\abstracttext
\end{abstract}


\section{Introduction}
\label{sec_intro}

Although most effort in cosmology has naturally been directed towards an understanding of the bright galaxies and clusters in the high-density peaks of the matter distribution in the Universe, the study of their counterparts in the cosmic web---the vast low-density regions known as cosmic voids----has recently gained significant importance for cosmology. 

As a consequence of their low matter content, voids become dominated by dark energy at early times, and so are sensitive to its nature \citep{lee09,lavaux12,bos12,pisani15}. The dynamics within voids can be accurately modelled by linear perturbation theory even on small scales \citep{Cai:2016a, Nadathur:2019a, Nadathur:2019b}, which provides a unique opportunity to measure the growth rate of structure through redshift-space distortions (RSD) in the distribution of galaxies around voids \citep[some examples of these studies in survey data include][]{Hamaus:2016,Hawken:2017,Achitouv:2017a,nadathur19}. A recent analysis by \citet{nadathur19} of the redshift-space void-galaxy correlation observed in the Baryon Oscillation Spectroscopic survey \citep[BOSS; ][]{Dawson:2013} showed that in combination with galaxy clustering, this method reduces the uncertainty in the measurement of cosmological distance scales by 50\% compared to previous results based on baryon acoustic oscillations \citep{Alam:2017}. An important ingredient for such studies is knowledge of the dark matter distribution within voids. This cannot be directly observed and so is currently calibrated from simulations but in principle could be inferred from measurement of the stacked gravitational lensing signal from voids \citep{Krause:2013}.

The matter distribution within voids is also interesting in its own right, as it has been shown to be sensitive to the sum of neutrino masses \citep[e.g.][]{massara15,banerjee16,kreisch19,zhang19} and to alternative theories of gravity \citep[e.g.][]{cai15,Barreira:2015,Falck:2017,Cautun:2018,Baker:2018,Paillas:2019}. This latter sensitivity is because voids constitute low-density environments within which the screening mechanisms of some modified gravity models do not apply. Several detections of the void lensing shear signal have been made in different data \citep{melchior14,clampitt15,sanchez16,Fang:2019}.

In addition to their lensing effect, voids also have a gravitational redshifting effect on photons traversing them, imprinting small secondary anisotropies on the cosmic microwave background (CMB) via the integrated Sachs-Wolfe (ISW) effect. An early high-significance observation of the void-ISW signal \citep{Granett:2008} was found to be strongly discrepant with predictions for the standard $\Lambda$ Cold Dark Matter ($\Lambda$CDM) cosmological model \citep{Nadathur:2012,Flender:2013}. This led to much subsequent work on the cross-correlation of voids with CMB temperature maps in newer data \citep[e.g.][]{Cai:2014,hotchkiss15_jubilee,Planck:2015ISW,Granett:2015,nadathur16,Kovacs:2017,Kovacs:2019}, although conclusions regarding the severity of the  discrepancy (if any) differ, and detection significances remain low. Less attention has been paid to the cross-correlation of voids with CMB lensing convergence maps---although  \citet{cai17} and recently \citet{vielzeuf19} have both reported $\sim3\sigma$ detections of CMB lensing by voids.

The cross-correlation between maps of the reconstructed CMB lensing convergence $\kappa$ and other tracers of the low-redshift large-scale structure has been the subject of much recent study \citep[e.g.][]{schmittfull18,SO19}. 
CMB lensing has been used to measure masses of dark matter haloes \citep[initial detections include][]{madhavacheril15, baxter15,plancksz15}.
Its correlation with cosmic filaments was also used to study the non-linearities in structure formation \citep{he18}. 
\citet{chantavat16} argued that the measurement of the CMB lensing by voids can be used as a probe of cosmological parameters. 

In this work, we use the full-sky reconstructed $\kappa$ map from the \planck{} 2018 data release \citep{planck18_lensing} and over 7000 voids extracted from the CMASS spectroscopic galaxy sample of the BOSS Data Release 12 catalogs to examine the CMB lensing imprint of voids. Our work uses similar data to that used by \citet{cai17}, who reported a $3.2\sigma$ detection of the void $\kappa$ signal, albeit with a slightly different void catalog and the latest \planck{} lensing reconstruction in place of the 2015 map. However, we introduce new improved methods for void stacking that greatly increase the detection sensitivity (\S\ref{sec_methods}). We calibrate theoretical expectations for the void lensing signal using mock void catalogs in a suite of 108 full-sky lensing simulations produced by \citet{takahashi17} in \S\ref{sec_calibration}. Relative to this expectation, in \S\ref{sec_results} we report measurement of a lensing amplitude of $A_L=1.10\pm0.21$ using an optimal matched filter technique and $A_L=0.97\pm0.19$ using an alternative Wiener filtering approach, representing detection of the void CMB lensing signal at significance levels of \mfcombinedsnr$\sigma$ and \wfcombinedsnr$\sigma$, respectively. We demonstrate that this detection is robust against thermal Sunyaev-Zel'dovich (tSZ) contaminations in the lensing reconstruction and other systematic effects. 

The improved measurement precision of the lensing convergence imprint of voids allows us to test the total matter distribution within these voids and to compare it to the distribution of visible galaxies. In \S\ref{sec_results}, we show that the void matter-overdensity profile $\delta(r)$ naively inferred from direct measurement of the galaxy density profile $\delta_g(r)$ and the assumption of a constant linear galaxy bias consistent with values obtained from galaxy clustering leads to a predicted lensing imprint that differs sharply from that obtained from calibration with the lensing simulations. This na\"ive bias model is also seen to be in disagreement with the measured $\kappa$ signal at $\sim 3\sigma$, predicting a lensing amplitude almost 40\% larger than that observed. We discuss why this discrepancy is expected due to selection effects arising from the fact that voids are selected as regions of low galaxy density, and show that it is consistent with previous results from simulations and data.

In \S\ref{sec_future_prospects} we forecast that the expected sensitivity for similar void lensing measurements improves by a factor of two or more using new CMB lensing data from current and next generation experiments \citep{henderson16, S4_DSR_19, SO19}
Finally, we summarize our results in \S\ref{sec_conclusion}.


\section{Datasets}
\label{sec_datasets}

\subsection{CMB lensing maps}
\label{sec_cmb_lensing}

We make use of the public CMB lensing convergence maps from the \planck{} 2018 data release \citep{planck18_lensing}.\footnote{Downloaded from \url{https://pla.esac.esa.int/\#cosmology}} Our fiducial analysis uses the map \texttt{COM\_Lensing\_4096\_R3.00} reconstructed using a minimum-variance (MV) quadratic estimator \citep{hu02} from a combination of foreground-cleaned \smica{} \citep{planck16_smica} CMB temperature and polarization maps, with the mean field subtracted and a conservative mask applied to galaxy clusters to reduce contamination from thermal Sunyaev-Zel'dovich (tSZ) contributions. 
As tSZ signals are known to be a potential contaminant for the CMB-lensing reconstruction \citep{vanengelen13, madhavacheril18}, and as \cite{alonso18} reported a detection of tSZ within voids, we also test for residual systematics in our measurement using a second convergence map (\texttt{COM\_Lensing-Szdeproj\_4096\_R3.00}) reconstructed from tSZ-deprojected \smica{} temperature data alone. 

For both maps, we use information from lensing modes $L \le \lmaxforanalysis$. Higher $L$ modes are highly noise-dominated for the \planck{} lensing reconstruction, and are in any case irrelevant for the void lensing signal of interest here, which varies on degree scales. We tested the use of an additional high-pass filter to restrict the multipole range to $8 \le L \le \lmaxforanalysis$ as used by \cite{planck18_lensing}, but found that it made negligible difference to the results obtained. Our default analysis presented below therefore does not exclude the largest scale modes $L < 8$.

\subsection{BOSS data and void catalog}
\label{sec_voids}

To construct the void catalog used we use the CMASS galaxy sample from BOSS \citep{Dawson:2013}) Data Release 12 galaxy catalogs \citep{Alam-DR11&12:2015}, which comprise the final data release of the third generation of the Sloan Digital Sky Survey (SDSS-III; \citealt{Eisenstein-BOSS:2011}). 
BOSS measured optical spectra for over 1.5 million targets covering nearly 10,000 deg$^2$ of the sky. 
The CMASS sample selection is based on colour-magnitude cuts designed to select massive galaxies in a narrow range of stellar mass with redshifts $0.43\le z\le0.7$ \citep{Reid-DR12:2016}. These galaxies are biased tracers of the matter distribution, with a bias of $b_\mathrm{CMASS}\sim2$ \citep{Alam:2017}. 
Voids from this CMASS sample have previously been used in a variety of works \citep[e.g.,][]{nadathur16_void_catalogue_paper, nadathur16, Hamaus:2016, cai17, nadathur19}.

We construct a void catalog from the CMASS data using the public \texttt{REVOLVER} void-finding code \citep{nadathur19, Nadathur:2019ascl}\footnote{Available from \url{https://github.com/seshnadathur/Revolver}} which is derived from the earlier \texttt{ZOBOV} algorithm \citep{neyrinck08}. \texttt{REVOLVER} estimates the local galaxy overdensity field from the discrete galaxy distribution using a Voronoi tessellation field estimator (VTFE) technique including additional corrections for the CMASS selection function and the survey angular completeness and masks using appropriate weights, as described in detail in \cite{nadathur14_void_catalogue_paper, nadathur16_void_catalogue_paper, nadathur19}. Locations of minima of this density field are identified as the sites of potential voids, the extents of which are determined by a watershed algorithm without pre-determined assumptions about void shapes. Following previous work \citep{nadathur16_void_catalogue_paper, nadathur15a}, we define each individual density basin as a distinct void, so that voids do not overlap. \texttt{REVOLVER} provides an option to remove RSD in the void positions through using density-field reconstruction prior to void finding \citep{Nadathur:2019b, nadathur19}, but this step has a negligibly small effect on the predicted lensing signal of voids, so is omitted here. Thus our void-finding procedure matches that previously used by \cite{nadathur16}.

The resulting catalog contains a total of \totalvoids{} voids, with a redshift distribution that is close to flat. Their low central density means that the lensing imprint of voids qualitatively corresponds to a de-magnification ($\kappa<0$) near the void center. The matter distribution around a typical void also shows an overdensity ($\delta>0$) around the void boundaries, caused by the pileup of matter evacuated from the center. These walls produce a ring feature of $\kappa>0$ around the central minimum. In this work, void centers are identified as the center of the largest sphere completely empty of galaxies that can be inscribed within the void, which is the best predictor of the location of the matter density minimum \citep{nadathur15a}. A commonly-used alternative choice is to define the void center as the weighted average position of the galaxies within it, or barycenter. This latter choice instead emphasises the high-density void walls and therefore the $\kappa>0$ ring, while smoothing out or even missing the central minimum. While this shape of the convergence profiles $\kappa(\theta)$ for voids is less intuitive, both dip- and ring-type imprints can be detected in lensing convergence maps, so the choice makes no practical difference to the detection sensitivity.

For each void, we calculate an average galaxy overdensity $\bar{\delta}_{g}$, defined as the volume-weighted average of the VTFE overdensity values in each of the Voronoi cells comprising the void \citep{nadathur17}, and an effective spherical radius $r_v$, defined as the radius of the sphere with the same volume as the (arbitrarily shaped) void. The nature of the watershed algorithm means that void extents are always defined to include the high-density regions in the separating walls, and as a result $\bar\delta_g$ is typically $\sim 0$ but can be either positive or negative (see the discussion in \citealt{nadathur15a}). Void sizes lie in the range $9.4\le r_v\le 111\ \rvunits$, with a well-defined maximum around the median value $r_v=41\ \rvunits$. The median void redshift is $z=0.55$, and the median angular scale subtended by spheres of the same $r_v$ would be $\sim1.6^\circ$.

From these values, for each void in the catalog we construct the dimensionless parameter
\begin{equation}
    \label{eq:lambda_v}
    \lambda_v \equiv \bar{\delta}_{g} \left(\frac{r_v}{1\ \rvunits}\right)^{1.2}\,,
\end{equation}
that \cite{nadathur17} empirically found to be tightly correlated with the void-density profiles and large-scale environments. 
In \S \ref{sec_calibration} and \S \ref{sec_methods} we discuss the scaling of the void lensing convergence profiles with $\lambda_v$ and how this informs our filtering templates.

\subsection{Lensing simulations}
\label{sec_takahashi_simulations}

\begin{figure*}[htb!]
\centering
\includegraphics[width=0.95\textwidth, keepaspectratio]{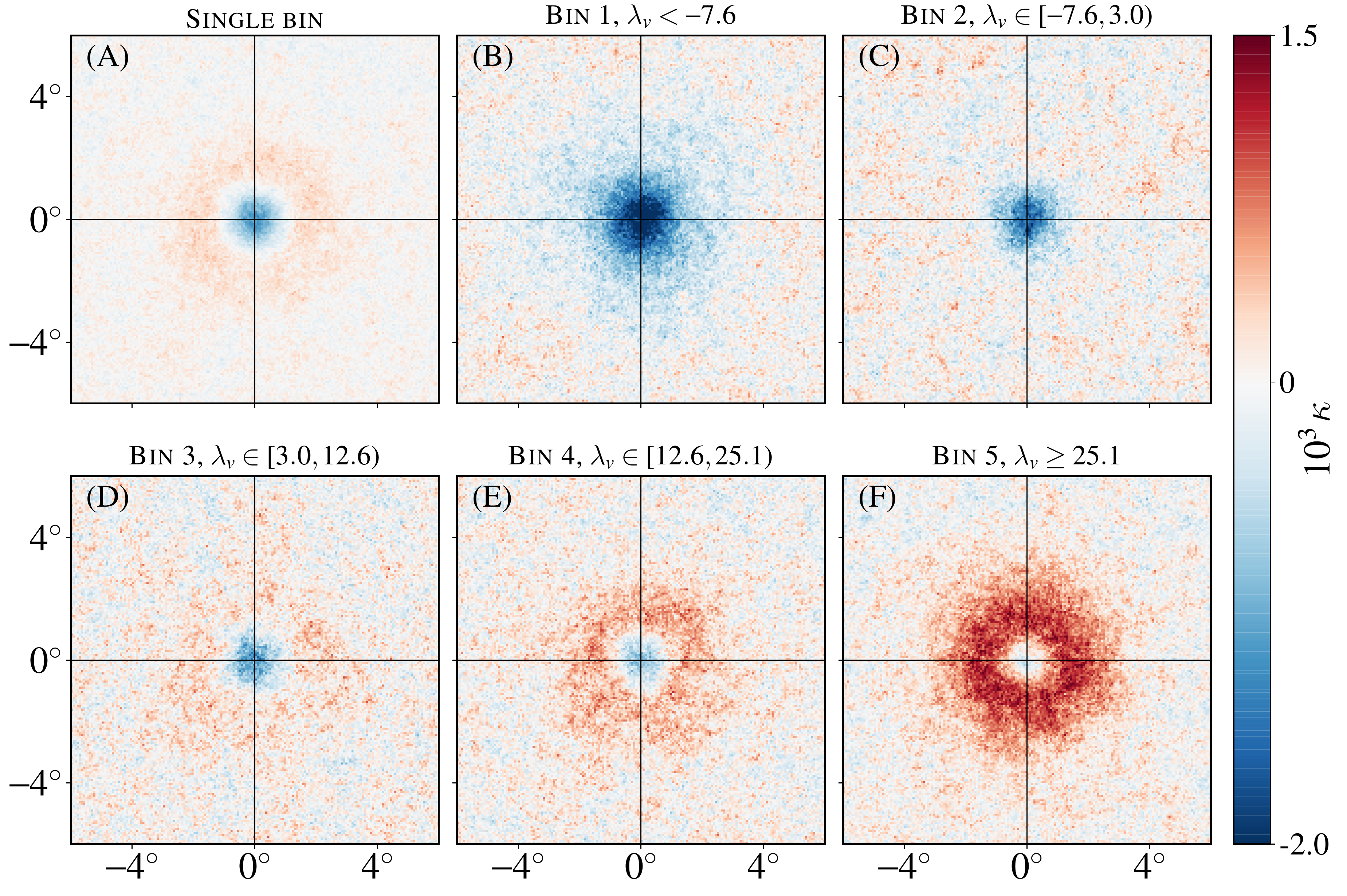}
\caption{
Stacked convergence ($10^{3} \kappa$) maps at void locations in the \takasims. 
Cutouts of size \boxsize{} centered at void positions were extracted from the lensing maps and averaged. Panel (A) shows the resultant stack for all voids in a single bin. Panels (B) through (F) show stacked signals for different sub-populations of voids, binned according to their $\lv$ values as indicated. Bin boundaries are chosen such that each bin contains roughly 1400 voids in each of the 108 realizations. The predominantly positive $\kappa$ signal in (F) is a projection effect caused by looking through high-density walls around the voids, which still correspond to genuine matter underdensities at their centers. 
}
\label{fig_lambda_5bins_stacks_takahashi_sims}
\end{figure*}

The contribution to the CMB lensing convergence profile from an isolated void with known spherically symmetric matter overdensity distribution $\delta(r)$ can be written as 
\begin{equation}
    \label{eq:kappa model}
    \kappa(\theta)=\frac{3\Omega_{m}H_0^2}{2c^2}\int \frac{\chi\left(\chi_s-\chi\right)}{\chi_s}\frac{\delta(\theta,\chi)}{a}\mathrm{d}\chi,
\end{equation}
where $\chi$ is the comoving radial coordinate and $\chi_s$ is the comoving distance to the last scattering surface. However, in general $\delta(r)$ is not known except from calibration with simulations, and voids are not completely isolated, so the effects of other structures along the line-of-sight need to be accounted for.

To make model predictions for the void lensing signal and to calibrate the optimal filters for application to data, we therefore make use of the public suite of full-sky lensing simulations described by \cite{takahashi17}\footnote{\url{http://cosmo.phys.hirosaki-u.ac.jp/takahasi/allsky\_raytracing/}}.
These consist of 108 realizations of full-sky lensing convergence and shear maps for all structures between redshifts $z=0.05$ to $5.3$, constructed from multiple $N$-body simulations in a flat $\Lambda$CDM cosmology run using \texttt{Gadget2} \citep{springel05}, with ray tracing performed using the public \texttt{GRayTrix} \citep{hamana15, shirasaki15} code. 
For this work, we use the maps corresponding to source redshift at the surface of last scattering, $z_s=1100$, labelled \texttt{zs=66}, in \healpix{} \citep{gorski05} format. 
We downsample the simulated maps from $N_\mathrm{side}=4096$ to $N_\mathrm{side}=2048$, corresponding to a pixel angular resolution of $1.^{\prime}7$. 

The cosmological parameters used for these simulations are based on the WMAP9 cosmology \citep{WMAP9}: $\Omega_m=0.279$, $\Omega_b=0.046$, $\Omega_\Lambda=0.721$, $n_s=0.97$, $h=0.7$, $\sigma_8=0.82$. These values unavoidably differ slightly from the \planck{} best-fit cosmology that is used elsewhere in this paper. We will assume that the effect of this on the calibration of our lensing templates is small and neglect it for the purposes of this work. Note that this is not an unreasonable approximation, as the most relevant parameter for determining the matter content of voids (and thus their lensing convergence $\kappa$) is $\sigma_8$ \citep{nadathur19}, and for the \takasims{} this is quite close to the \planck{} value $\sigma_8=0.81$ \citep{Planck:2018params}.

Halo catalogs on the lightcone are provided with each of these simulations. In the redshift range $0.4\lesssim z\lesssim 0.75$ of interest to us, the minimum halo mass resolution is $2\times10^{12}\;h^{-1}M_\odot$. From these halo catalogs we create galaxy mocks using the Halo Occupation Distribution (HOD) model of \citep{Zheng:2007}, with parameters as specified by \cite{Manera:2013} in order to match the clustering properties of CMASS galaxies. We apply the BOSS survey footprint and angular and radial selection functions in order to match the CMASS sample as closely as possible. To each mock catalog we then apply the same void-finding procedure described in \S \ref{sec_voids} as used for the BOSS data, to obtain 108 mock void catalogs, each consisting of $\sim 7000$ voids.

\subsection{MD-Patchy mock void catalogs}
\label{sec_patchy}

In order to reliably estimate the covariance matrix for the lensing measurements, it is desirable to use as large a sample of mocks as possible. The Takahashi simulations provide only 108 realizations, so we use voids from a suite of 2048 MD-Patchy mock galaxy catalogs created for the BOSS DR12 data release \citep{Kitaura-DR12mocks:2016} instead.\footnote{Alternatively, one could keep the BOSS void catalog fixed and repeat the stacking on the public \planck{} lensing simulations. But only 300 lensing realizations are available, which is also small relative to the size of the covariance matrix to be estimated (\S\ref{sec_fitting}).}
These mocks are created using the fast {\small PATCHY} algorithm, based on approximate simulations using augmented Lagrangian perturbation theory (ALPT,  \citealt{Kitaura:2013}). 
Mock galaxies are painted in dark matter haloes using a halo abundance matching algorithm \citep{Rodriguez-Torres:2015} trained on a reference full {\it N}-body simulation from the Big MultiDark suite \citep{Klypin:2016}. 
The mocks were designed and validated to match the clustering of the CMASS sample and to reproduce the selection functions and observational systematics. Note that the MD-Patchy mocks are used only for covariance matrix estimation and not template calibration, as they do not have associated CMB lensing simulations.

We run \texttt{REVOLVER} on each of these MD-Patchy mocks in the same way as for the CMASS data sample, thus obtaining 2048 mock void catalogs that statistically closely match the BOSS voids. Similar mock void catalogs have been used for covariance estimation for void measurements by \cite{nadathur16, nadathur19}. These MD-Patchy voids have the same clustering properties as the BOSS voids and occupy the same section of the \planck{} sky defined by the BOSS footprint, but are uncorrelated with real structures and thus with the \planck{} $\kappa$ map. This is expected to be a sufficient approximation for error bar calculation; the CMB-lensing and void fields should only have a modest correlation coefficient, so that the correlated cosmic variance contribution to the errors is expected to be negligible.

\section{Void lensing in simulation}
\label{sec_calibration}

We start by analysing the void lensing convergence signal seen in the \takasims{} from \S \ref{sec_takahashi_simulations} in order to calibrate theoretical expectations for the void lensing profile $\kappa(\theta)$. 
Panel (A) in Figure~\ref{fig_lambda_5bins_stacks_takahashi_sims} shows the stacked average $\kappa$ signal around void lines of sight in the simulations, constructed by stacking equal \boxsize{} patches cut out from the full-sky $\kappa$ maps centered at void positions. This stack contains all the voids from all of the realizations of the simulations, and shows qualitative features in accord with intuition: a small central region with $\kappa<0$, (i.e., a de-magnification due to the central void underdensity), surrounded by a larger but less-pronounced positive convergence ring corresponding to the location of the overdensity at the void boundary, caused by the pile-up of matter evacuated from the void center.

Several authors \citep[e.g.,][]{hotchkiss15_jubilee, cai17, Kovacs:2017, Kovacs:2019, vielzeuf19} advocate rescaling the angular sizes of each cutout based on the angular scale corresponding to the individual void radius $r_v$ before stacking. Under the assumption that the angular sizes of the void lensing imprints of interest scale self-similarly with the void radius $r_v$, such a rescaling procedure would maximise the signal amplitude. However, \cite{nadathur17} showed that the shapes of void lensing-convergence profiles are much more strongly correlated with the combination of void size and density encapsulated in parameter $\lambda_v$ defined by Eq. \ref{eq:lambda_v}, than with $r_v$ alone. 

We therefore bin the simulation void samples into $N_\mathrm{bin}=5$ bins of $\lambda_v$, and perform the stacking separately in each bin, shown in panels (B) through (F) in Figure \ref{fig_lambda_5bins_stacks_takahashi_sims}. The bin boundaries were chosen based on quintiles of the $\lambda_v$ distribution for the BOSS voids. Each realization of the \takasims{} then contains $\sim1400$ voids in each bin.\footnote{We chose $N_\mathrm{bin}=5$ for convenience. In principle, $N_\mathrm{bin}$ should be as large as possible provided uncertainties in the template calibration in each bin remain negligible compared to data uncertainties. However, tests for $N_\mathrm{bin}>5$ showed no significant improvement in the expected \snr{} given \planck{} noise levels.} The void lensing signal shows an extremely strong dependence on $\lambda_v$. Negative values of $\lambda_v$ (bins 1 and 2) correspond to voids embedded within low-density regions, producing $\kappa\le0$ everywhere. Voids with large positive $\lambda_v$ (bin 5) correspond to local minima within larger-scale overdensities, producing a very pronounced $\kappa>0$ convergence ring. 
In other words, small (i.e., large negative) values of $\lambda_v$ correspond to R-type voids \citep{ceccarelli13, paz13, ruiz18} or \emph{voids-in-voids} \citep{Sheth:2004} within larger-scale underdensities, whereas large positive values of $\lambda_v$ correspond to S-type voids or \emph{voids-in-clouds}, local density minima sitting within a larger-scale overdensity.
The mean void sizes in bins 1 to 5 are $\bar{r}_v=50.5$, 41.3, 39.6, 38.2 and 37.8 $\rvunits$, respectively, but these values do not correspond to the angular scales subtended by the lensing imprints. The advantage of separating voids by $\lambda_v$ is clear by comparison to panel (A): if voids of different $\lambda_v$ are stacked together, the resultant $\kappa$ signal averages out to a value closer to zero and is consequently harder to detect. Note that the stack for voids in bin 5, with the largest $\lambda_v$ values, shows primarily positive convergence $\kappa>0$ as might be expected from an overdensity. However, this is a projection effect caused by looking through the void walls: these voids do still correspond to genuine underdensities in the matter distribution, with on average $\delta\lesssim-0.4$ at their centers (for instance, see the $\delta(r)$ profiles in Fig. 6 of \citealt{nadathur17}).

\begin{figure}
\centering
\hspace{-5mm}
\includegraphics[width=0.48\textwidth,keepaspectratio]{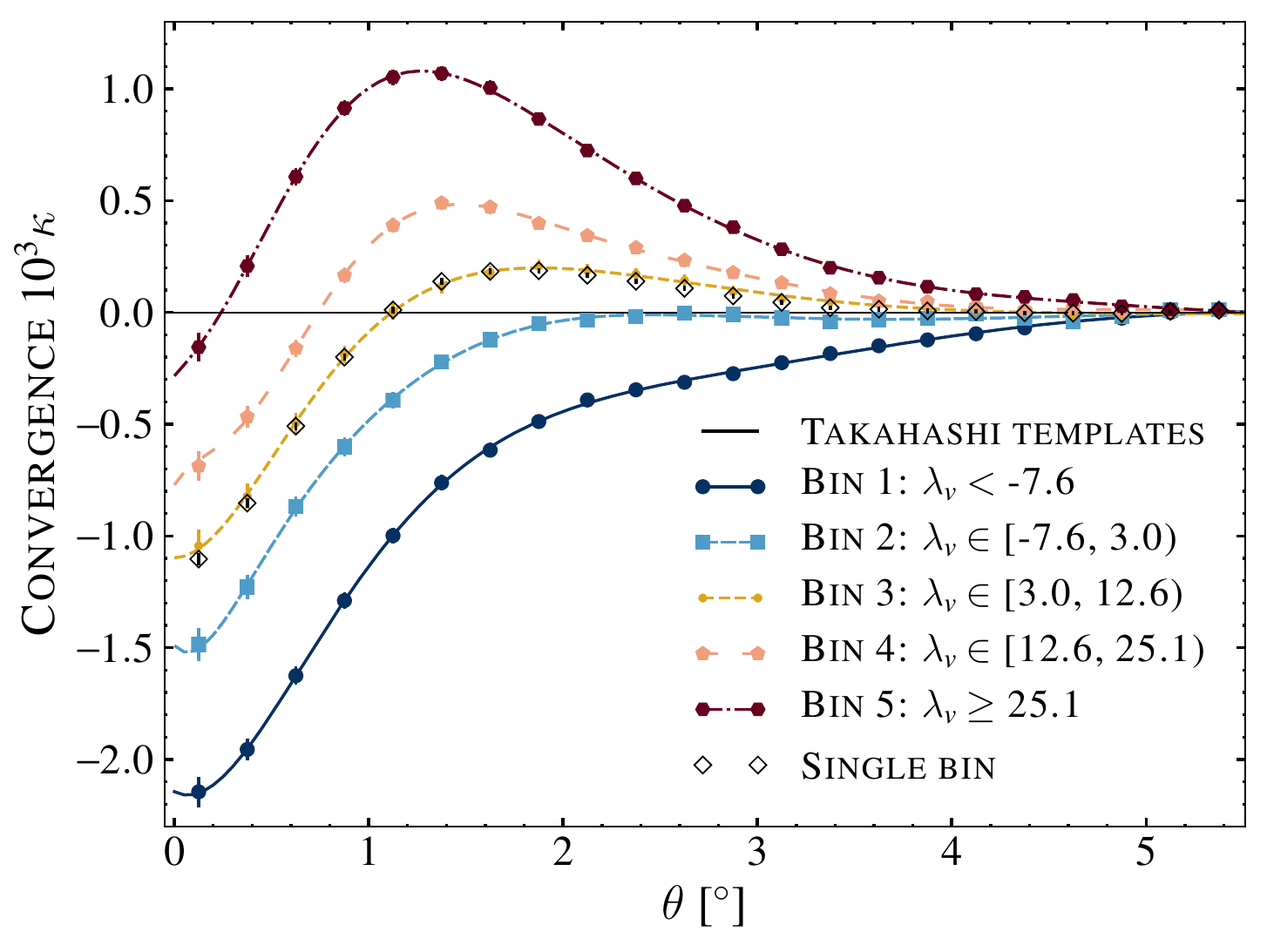}
\caption{Stacked radial convergence profiles $\kappa(\theta)$ for voids in the \takasims. 
Data points show the mean values for voids in the respective bins averaged over all 108 realizations, while error bars indicate the $68\%$ uncertainties in the mean of a single realization containing $\sim7000$ voids, or $\sim 1400$ in each $\lv$ bin. The curves are polynomial template fits to the simulation results, which we use to create the matched filters and model the \planck{} data.
}
\label{fig_lambda_5bins_radprf_takahashi_sims}
\end{figure}

From these stacks, we measure azimuthally-averaged 1D convergence profiles $\kappa(\theta)$ for each $\lambda_v$ bin, shown in Figure~\ref{fig_lambda_5bins_radprf_takahashi_sims}. Data points represent the mean convergence value averaged over all voids in each $\lv$ bin over all 108 simulation realizations. Error bars represent the $68\%$ C.L. uncertainty in the mean convergence for voids in an individual realization; this represents the theoretical uncertainty in the mean signal for a CMASS-like void sample. 
For each $\lambda_v$ bin, we fit a polynomial function to the data points and use this to define a template profile $\kappa_\mathrm{template}(\theta;\lv)$ for each stack. 
The curves in Figure~\ref{fig_lambda_5bins_radprf_takahashi_sims} correspond to the templates using polynomial fits.
Note that these convergence profiles represent the pixel-space void CMB-lensing cross-correlation signals.

\section{Methods}
\label{sec_methods}

\subsection{Filtering the lensing map}
\label{sec_filters}

\begin{figure}
\centering
\includegraphics[width=0.45\textwidth]{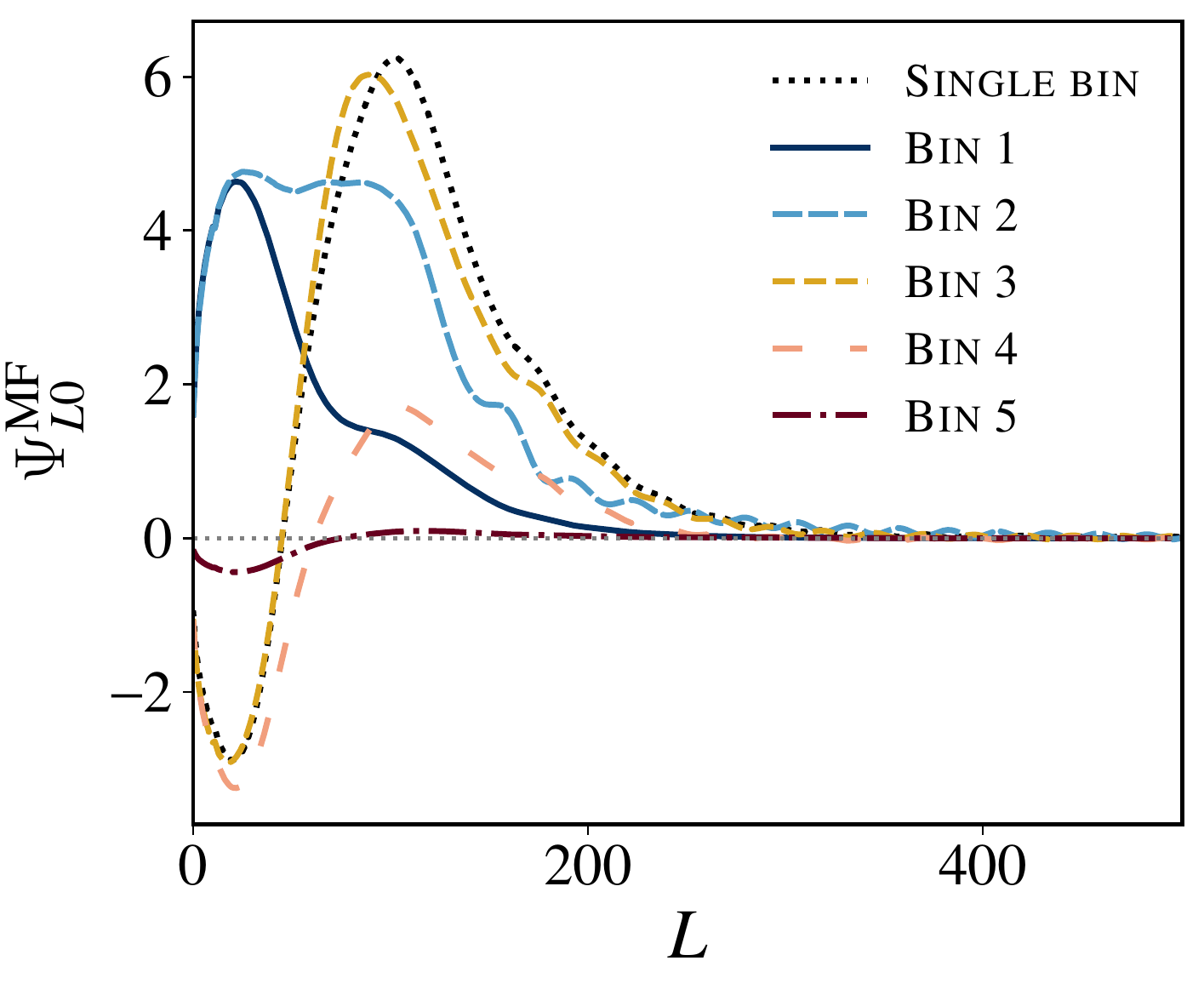}
\caption{
Spherical harmonic coefficients $\Psi^\mathrm{MF}_{L0}$ for the optimal matched filters designed for the detection of the template lensing profiles in each void $\lambda_v$ bin, shown in Figure~\ref{fig_lambda_5bins_radprf_takahashi_sims}. The filters amplify large-scale power and strongly suppress the small scales where voids do not contribute lensing information, with essentially no power at $L>300$ in any bin.
}
\label{fig_MF_filter_responses}
\end{figure}

\begin{figure*}
\centering
\includegraphics[width=0.45\textwidth]{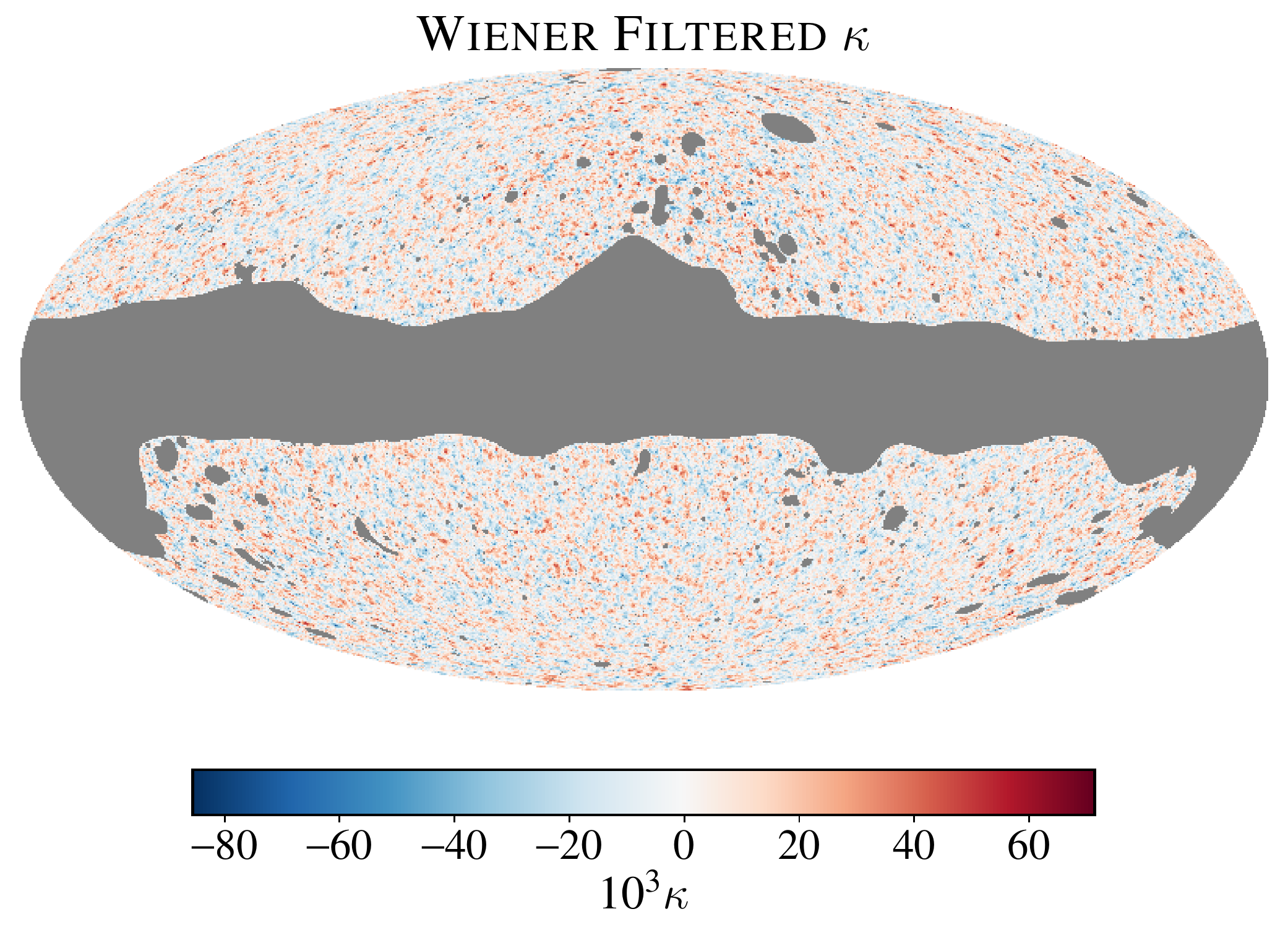}
\includegraphics[width=0.45\textwidth]{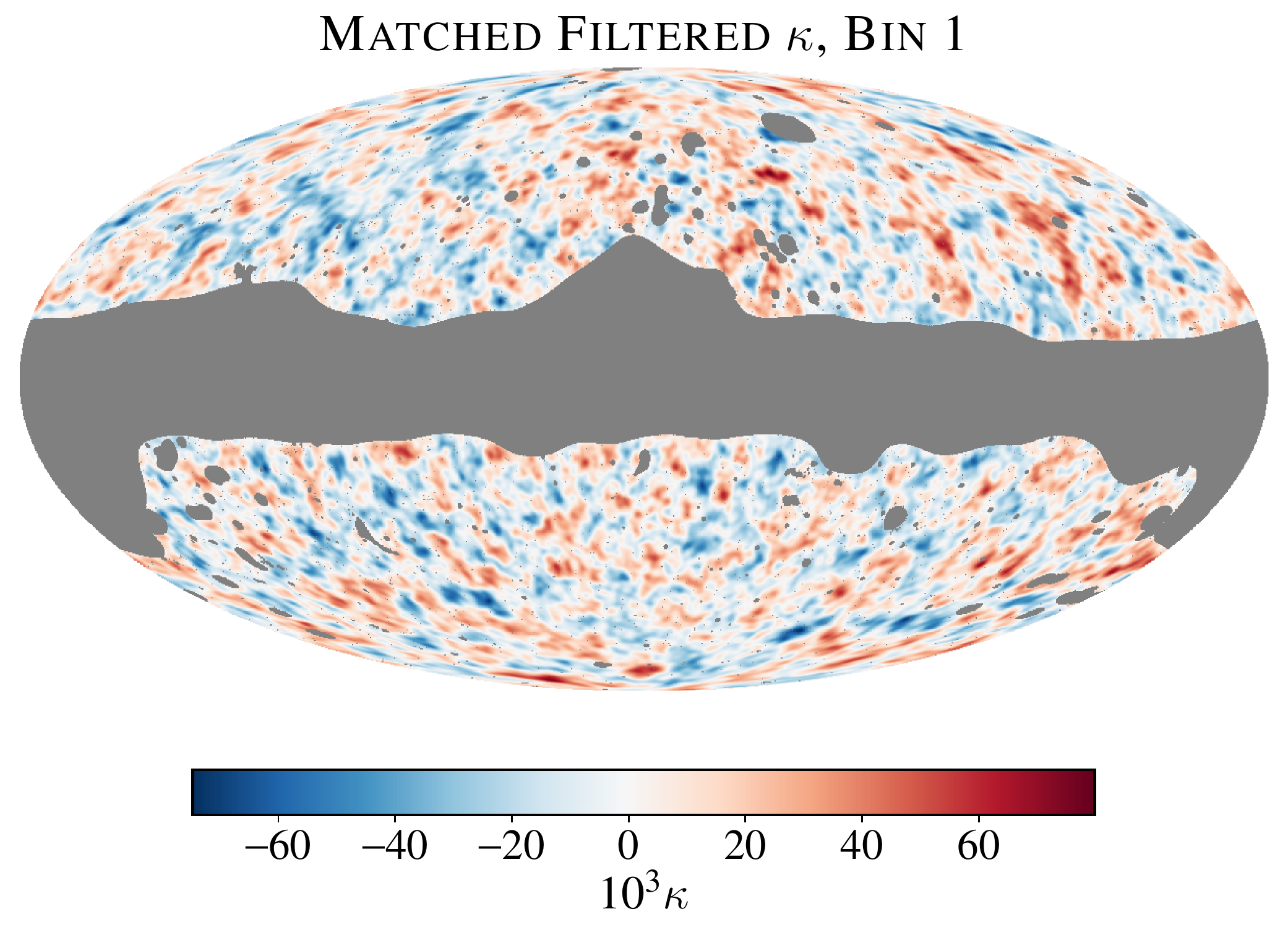}\\
\includegraphics[width=0.45\textwidth]{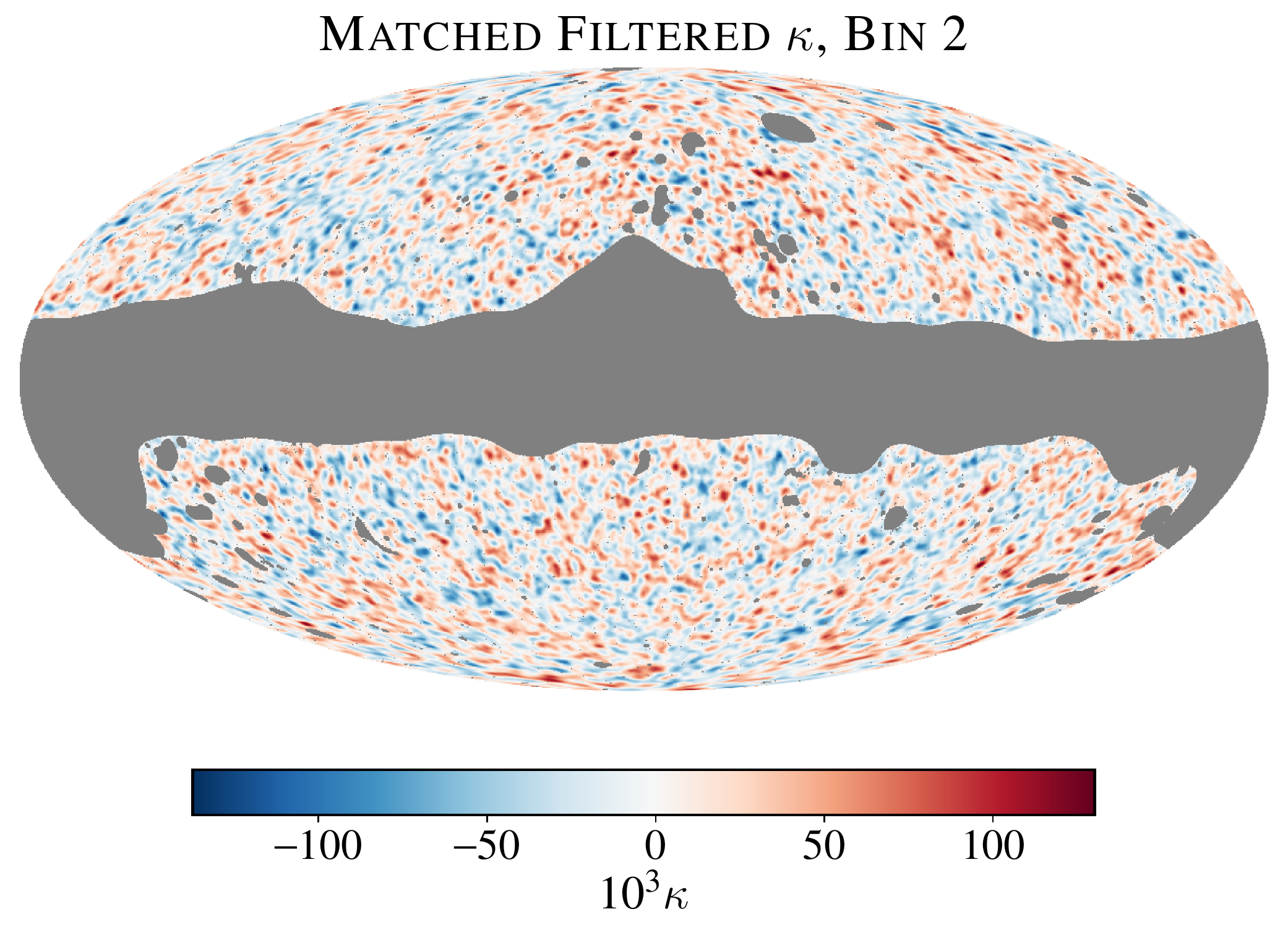}
\includegraphics[width=0.45\textwidth]{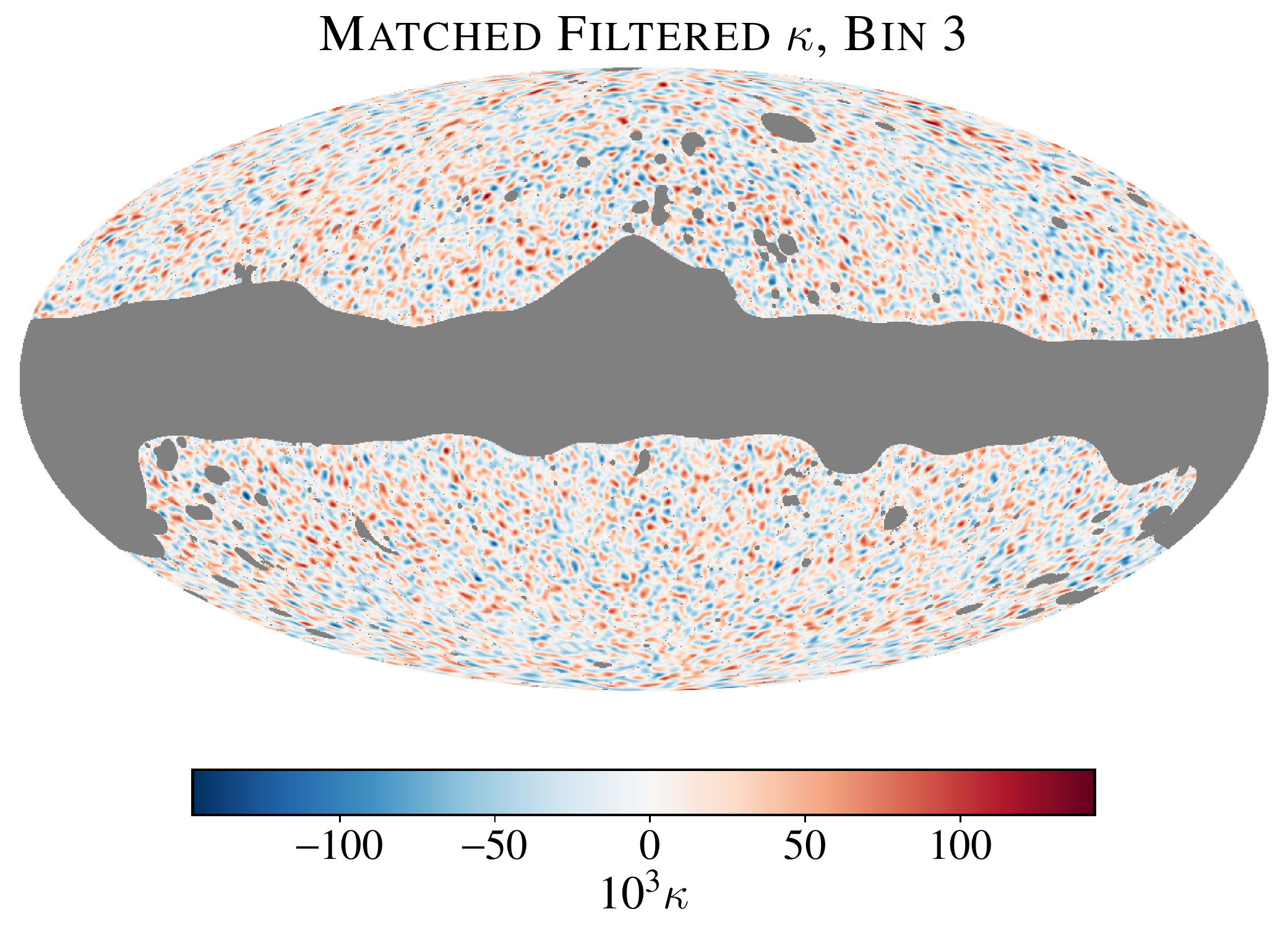}\\
\includegraphics[width=0.45\textwidth]{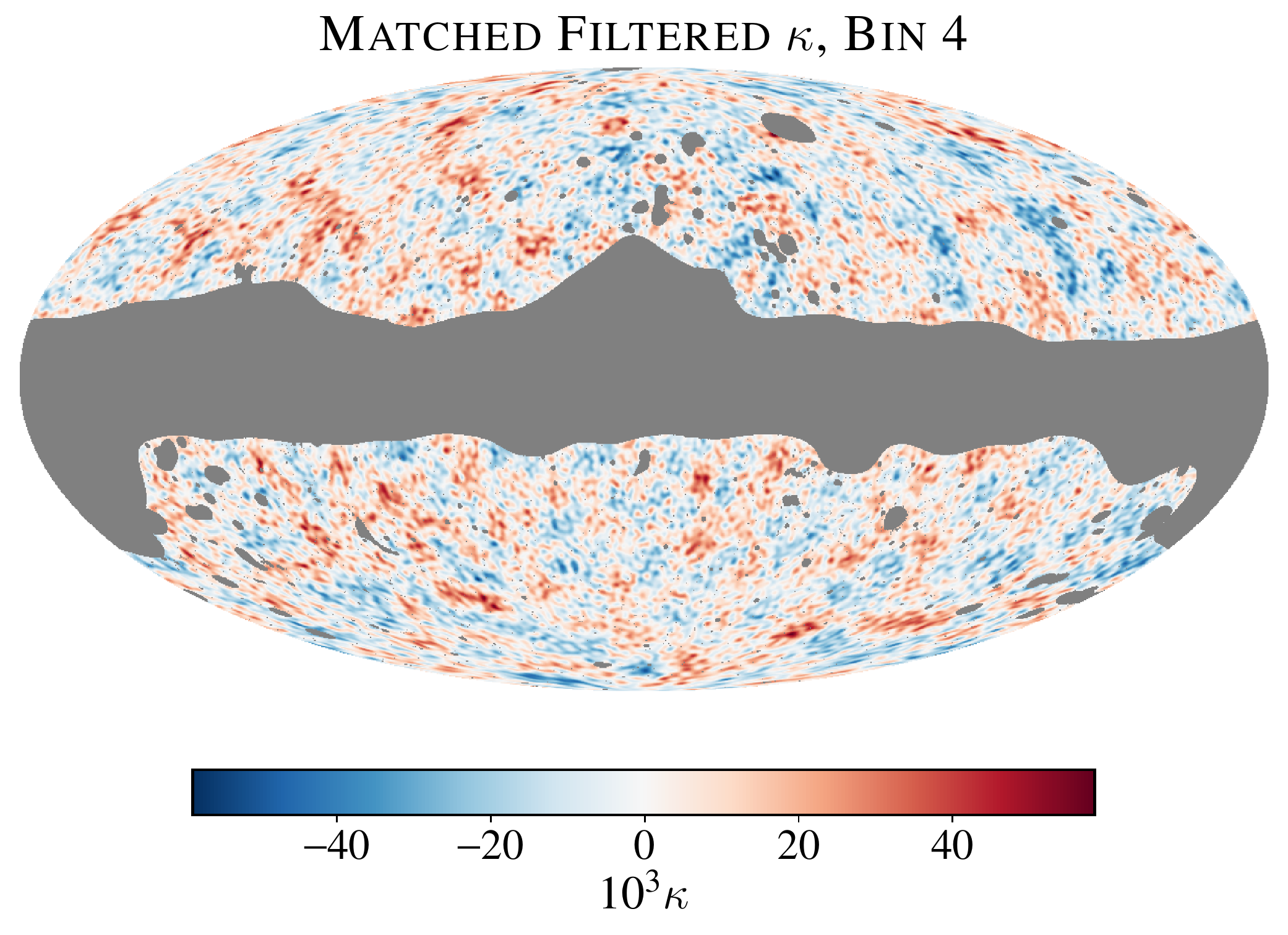}
\includegraphics[width=0.45\textwidth]{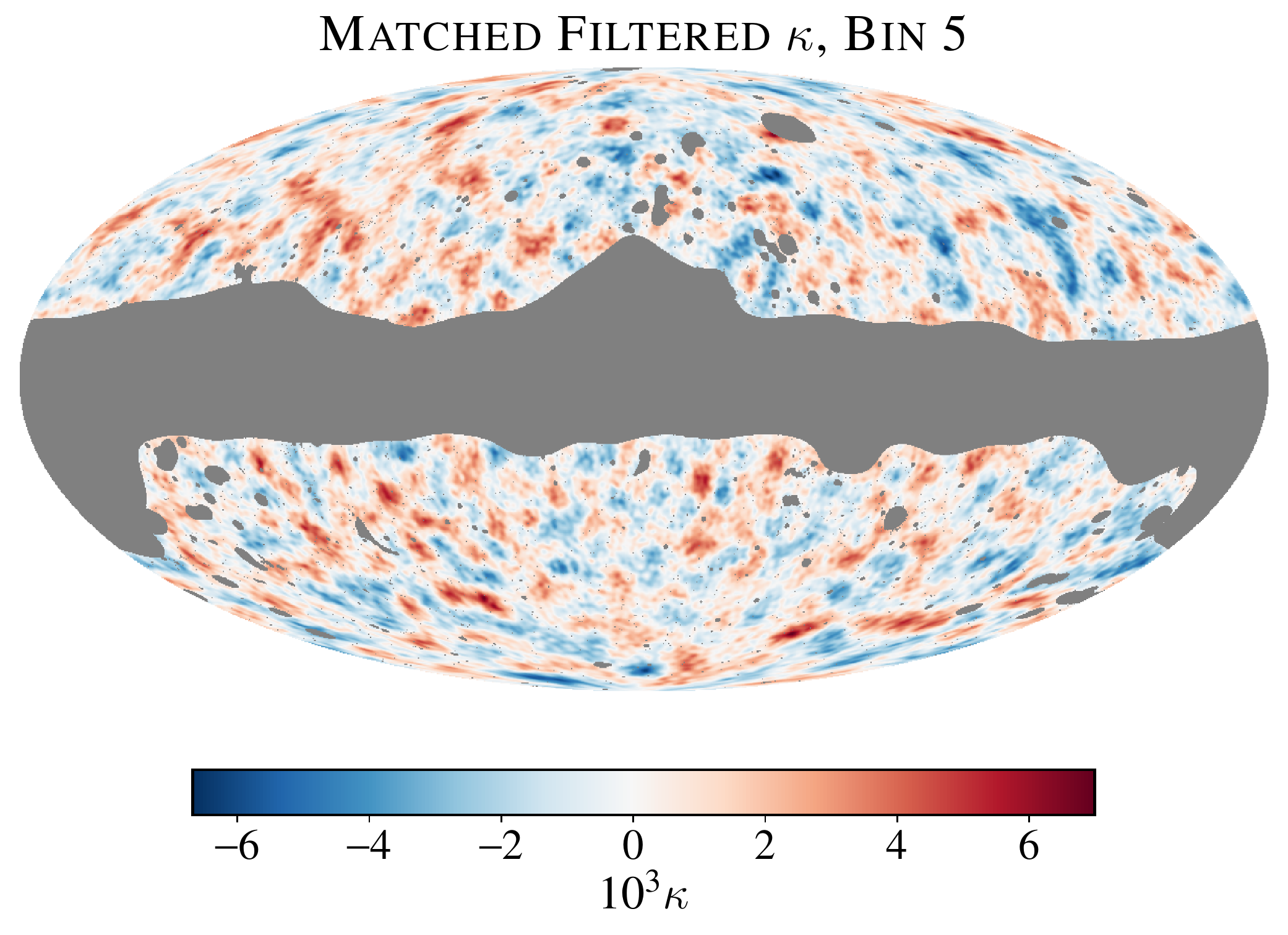}
\caption{
Filtered versions of the \planck{} lensing convergence map. We show the Wiener-filtered map, and the map convolved with each of the optimal matched filters for the $\lambda_v$ bin templates shown in Figure~\ref{fig_MF_filter_responses}.
}
\label{fig_filtered_Planck_maps}
\end{figure*}

Two significant sources of noise affect the measurement of the void-lensing cross-correlation: the lensing reconstruction noise in the \planck{} $\kappa$ map, and the contribution to $\kappa$ from uncorrelated structures along the line of sight. Both these contributions are orders of magnitude larger than the signal of interest, so the lensing imprint of individual voids is undetectable. This situation is improved by stacking many voids together, especially when the stacks are separated into $\lambda_v$ bins as discussed in \S\ref{sec_calibration} above. In addition to this, the application of well-chosen filters to the $\kappa$ map before stacking can improve the detection sensitivity.

In this work we follow two different filtering approaches and check that the results obtained from both are in good agreement. The first approach is based on applying a Wiener filter to the \planck{} reconstructed $\kappa$ map in order to down-weight the noise-dominated modes. In spherical harmonic space, the action of the Wiener filter is described by
\begin{equation}
    \label{eq_wiener}
    \kappa^\mathrm{WF}_{LM} = \frac{C_L^{\kappa\kappa}}{C_L^{\kappa\kappa} + N_L^{\kappa\kappa}}\kappa_{LM}\,,
\end{equation}
where $C_L^{\kappa\kappa}$ and $N_L^{\kappa\kappa}$ are respectively the lensing and noise power spectra for the \planck{} lensing map \citep{planck18_lensing}. 
The stacking analysis described below is then performed on patches extracted from this Wiener-filtered map. Note that the design of the Wiener filter in Eq.~\ref{eq_wiener} requires knowledge of the lensing reconstruction noise and the overall lensing power for all structures along the line of sight, but does not require knowledge of the expected void lensing imprints of interest obtained from simulation in the previous section. 
Thus by construction this Wiener filter does not reduce the variance sourced by other structures along the line of sight. 
This filter could in principle also be modified based on the lensing templates. However, this is already achieved by the optimal matched filter method described below and keeping the Wiener filtering independent of the simulation templates serves a useful cross-check of that approach.

The second approach we follow is to design optimal matched filters based on the simulation templates obtained in \S\ref{sec_calibration}. To describe the construction of the matched filters, we first represent the total convergence field at a point $\boldsymbol{\theta}$ in the vicinity of the position $\boldsymbol{\theta}_0$ of a void as 
\begin{equation}
    \label{eq:kappa split}
    \kappa(\boldsymbol{\theta}) = \kappa_\mathrm{template}(\vert\boldsymbol{\theta}-\boldsymbol{\theta}_0\vert; \lambda_v) + n(\boldsymbol{\theta})\,,
\end{equation}
where $n$ represents a generalised noise term that includes all features in the convergence map other than the desired void signal, and $\kappa_\mathrm{template}$ describes the appropriate void lensing template profile obtained from the \takasims{}. We further decompose this template profile as 
\begin{eqnarray}
    \label{eq_template_SH}
    \kappa_\mathrm{template}(\theta;\lambda_v) &=& \kappa_0(\lambda_v)k(\theta;\lambda_v)\\ \nonumber
    &=& \kappa_0(\lambda_v)\sum_{L=0}^{\infty}k_{L 0}(\lambda_v)Y_L^0(\cos\theta)\,,
\end{eqnarray}
where we have split it into an amplitude term $\kappa_0(\lambda_v)\equiv\kappa_\mathrm{template}(0;\lambda_v)$, and a normalised shape function $k(\theta)$ defined by the spherical harmonic coefficients $k_{L0}$. 

Given this decomposition and the assumption that the noise term is homogeneous and isotropic with zero mean, the spherical harmonic coefficients of the optimal matched filter are uniquely determined \citep{Schaefer:2006, McEwen:2008} to be:
\begin{equation}
    \label{eq_matched_filter}
    \Psi^\mathrm{MF}_{L0}(\lambda_v) = \alpha\frac{k_{L0}(\lambda_v)}{C_L^{N,tot}}\,,
\end{equation}
where 
\begin{equation}
    \label{eq_alpha}
    \alpha^{-1}\equiv\sum_{L=0}^{\infty}\frac{k_{L0}^2(\lambda_v)}{C_L^{N,\mathrm{tot}}}\,,
\end{equation}
and $C_L^{N,\mathrm{tot}} = C_L^{\kappa \kappa} + N_L^{\kappa \kappa}$ is the total power spectrum of the noise field. Figure~\ref{fig_MF_filter_responses} shows the optimal matched filters in each $\lambda_v$ bin, designed for the template profiles obtained in \S \ref{sec_calibration}. For comparison, we also show the appropriate matched filter for the stack of all voids together in single bin as black dotted line; this is naturally very close to that for the central $\lambda_v$ bin. Note that the templates corresponding to bin 1 and bin 2 in Figure~\ref{fig_lambda_5bins_radprf_takahashi_sims} do not change sign, and therefore the matched filters for these bins do not do so either. For all other bins, a clear crossover point is seen, leading to filter profiles that are partially or fully compensated.

For the matched filter defined in each $\lambda_v$ bin, the filtered lensing map $\kappa^\mathrm{MF}$ is a convolution for the filter with the original map,
\begin{equation}
    \label{eq:filter convolution}
    \kappa^\mathrm{MF}(\boldsymbol{\beta}) = \int \mathrm{d}\Omega\, \kappa(\boldsymbol{\theta})\Psi^\mathrm{MF}\left(|\boldsymbol{\theta}-\boldsymbol{\beta}|\right)\,,
\end{equation}
which can be written in spherical harmonic space as \citep{Schaefer:2006},
\begin{equation}
    \label{eq:filter SH}
     \kappa^\mathrm{MF}_{LM} = \sqrt{\frac{4\pi}{2L + 1}}\kappa_{LM}\Psi^\mathrm{MF}_{L 0} \,.
\end{equation}
The matched-filtered maps for each of the five $\lambda_v$ bins, together with the Wiener-filtered map, are shown in Figure \ref{fig_filtered_Planck_maps}. 

The construction of the matched filter ensures that the expectation value of the filtered field at void locations is 
\begin{equation}
    \label{eq:MF expectation}
    \langle\kappa^\mathrm{MF}(\boldsymbol{0};\lambda_v)\rangle = \kappa_0(\lambda_v)\,,
\end{equation}
meaning that the filter is \emph{unbiased}, and the variance of the filtered field at this location, $\sigma^2_\mathrm{MF}(\boldsymbol{0};\lambda_v)=\sum_{L=0}^\infty C_L^{N,\mathrm{tot}}\vert\Psi^\mathrm{MF}_{L0}\vert^2$, is minimized. The power of the optimal matched filter can be quantified in terms of the maximum detection level \citep{McEwen:2008} for a single isolated void,
\begin{equation}
    \label{eq_Gamma}
    \Gamma_{1v}(\lambda_v) \equiv \frac{\langle\kappa^\mathrm{MF}(\boldsymbol{0};\lambda_v)\rangle}{\sigma_\mathrm{MF}(\boldsymbol{0};\lambda_v)} = \alpha^{-1/2}\kappa_0(\lambda_v)\,.
\end{equation}
From this we also calculate a related quantity, $\Gamma_\textsc{BOSS}$, which is the corresponding maximum detection level for stacks containing as many voids in each bin as are present in the BOSS void catalog. This is calculated by simply dividing the noise by a factor of $\sqrt{N_v}$, and so assumes the void positions are independent of each other and that their profiles do not overlap. It therefore represents an upper bound on the true achievable detection significance. The $\Gamma_{1v}$ and $\Gamma_\textsc{BOSS}$ values obtained are given in Table \ref{table_Gamma}. Comparison with the corresponding values for the stack of all voids together again highlights the advantage of the $\lambda_v$-binning strategy employed here. It also shows that the two extreme $\lambda_v$ bins present by far the most easily detectable lensing signals, as expected from Figures \ref{fig_lambda_5bins_stacks_takahashi_sims} and \ref{fig_lambda_5bins_radprf_takahashi_sims}.

\begin{deluxetable}{lcccc}
\tablecaption{Expected maximum detection levels under optimal matched filters for a single void, and for the BOSS DR12 void catalog.}
\label{table_Gamma}
\tablehead{
\colhead{Void stack} & \colhead{$\lv$} &  \colhead{BOSS $N_v$} & \colhead{$\Gamma_{1v}$} & \colhead{$\Gamma_\textsc{BOSS}$}}
\startdata
{\string Single bin} & [-60.8, 159.8] & 7378 & 0.033 & 2.80  \\
{\string Bin 1} & [-60.8, -7.6) & 1478 & 0.105 & 4.04  \\
{\string Bin 2} & [-7.6, 3.0) & 1475 & 0.047 & 1.80  \\
{\string Bin 3} & [3.0, 12.6) & 1473 & 0.033 & 1.27  \\
{\string Bin 4} & [12.6, 25.1) & 1477 & 0.050 & 1.91 \\
{\string Bin 5} &[25.1, 159.8]  & 1475 & 0.110 & 4.22 \\
\enddata
\end{deluxetable}

\subsection{Detecting the void lensing signal}
\label{sec_fitting}

We extract mean-subtracted \boxsize{} patches centered at the location of each void $i$ in the catalog from the Wiener-filtered \planck{} $\kappa$ map, using which we measure the azimuthally-averaged profile $\hat{\kappa}_{i}^{\rm WF}(\theta)$ in 20 bins of \mbox{$\Delta \theta = \thetabinsize$}, out to a maximum $\theta_\mathrm{max}=5^\circ$. The final Wiener-filtered stacked profile is then obtained as 
\begin{equation}
    \hat{\kappa}^{\rm WF}(\theta) = \frac{\sum_i  w_i  \hat{\kappa}_{i}^{\rm WF}(\theta)}{\sum_i w_i}\,.
    \label{eq_stack}
\end{equation}
This measurement is repeated for voids in each $\lambda_v$ bin. 

The theoretical model to which we compare this observed quantity is
\begin{equation}
    \label{eq_AL}
    \kappa^\mathrm{th}(\theta) = \Al \kappa^\mathrm{WF}_{\rm template}(\theta)\,,
\end{equation}
where $\kappa^\mathrm{WF}_{\rm template}(\theta)$ denotes the template profile for that $\lambda_v$ bin calibrated from the simulations after convolution with the Wiener filter from Eq. \ref{eq_wiener}, and $\Al$ is a free fit parameter representing the lensing amplitude relative to that in the simulation templates. 

For the matched-filter analysis the filter design itself accounts for the lensing profile, so the measured quantity of interest is only the convergence at the void center location, $\kappa^\mathrm{MF}(\theta = 0)$. The stacked measurement in this case is
\begin{equation}
    \hat{\kappa}^{\rm MF}(0) = \frac{\sum_i  w_i \left[ (\hat{\kappa}_{i}^{\rm MF}(0) - \left< \hat{\kappa}_{r}^{\rm MF} \right> \right]}{\sum_i w_i}\,.
    \label{eq_MF_stack}
\end{equation}
Here $\left< \hat{\kappa}_{r}^{\rm MF} \right>$ is the mean value for the BOSS survey footprint. This is estimated from the mean convergence value in the filtered \planck{} map at the locations of voids in all the MD-Patchy mock catalogues, which cover the same footprint but are uncorrelated with true lensing signals. This mean subtraction is necessary because the matched filter design means that the filtered maps shown in Figure \ref{fig_filtered_Planck_maps} contain significant power on scales that are large compared to the sky fraction covered by BOSS (around 23\%). The theory model in this case is simply 
\begin{equation}
    \label{eq_MF_AL}
    \kappa^\mathrm{th}(0) = \Al \kappa_0,
\end{equation} 
where as in Eq. \ref{eq_AL} above, we have suppressed explicit $\lambda_v$-dependence for simplicity. 
For the purposes of the current work we apply a uniform void weighting, $w_i=w=1$ when calculating both Eqs. \ref{eq_stack} and \ref{eq_MF_stack}.

The covariance matrix $\mathbf{\hat{C}}$ for each measurement above is estimated using the $N_m=2048$ MD-Patchy mock void catalogs as
\begin{eqnarray}
    \label{eq_cov_matrix}
    \mathbf{\hat{C}} = \frac{1}{N_{m}-1}\sum\limits_{i = 1}^{N_{m}} \left(\boldsymbol{\hat{\kappa}_i} - \left< \boldsymbol{\kappa} \right> \right) \left(\boldsymbol{\hat{\kappa}_i} - \left< \boldsymbol{\kappa}\right> \right)^{T}\,.
\end{eqnarray} 
Here $\boldsymbol{\hat{\kappa}_{i}}$ is the data vector obtained from repeating the measurement in Eq.~\ref{eq_stack} or \ref{eq_MF_stack} using the filtered \planck{} map but with the void catalog obtained from the $i$th MD-Patchy mock. All $\lambda_v$ bin measurements are concatenated in the data vector, so that each vector $\boldsymbol{\hat{\kappa}_{i}}$ has dimensions $p=5\times20=100$ for the Wiener-filtered analysis where we measure the profile as a function of $\theta$, and $p=5$ for the matched-filter case. The covariance matrix therefore has corresponding dimensions $100\times100$ or $5\times5$ for the two analyses. For the Wiener-filtered stacking, the covariance matrix has high off-diagonal correlations between radial $\theta$ bins.

Given these ingredients, for any value of the model lensing amplitude $\Al$, we can calculate
\begin{equation}
    \label{eq_chi_sq}
    \chi^2 = \sum_{ij} \left(\hat{\kappa}_i - \kappa^\mathrm{th}_i \right) \mathbf{\hat{C}}_{ij}^{-1} \left(\hat{\kappa}_j - \kappa^\mathrm{th}_j \right)\,,
\end{equation}
where indices $i,j$ run over the $p=100$ ($p=5$) dimensions of the data and theory vectors defined by Eqs. \ref{eq_stack} and \ref{eq_AL} (Eqs. \ref{eq_MF_stack} and \ref{eq_MF_AL}) in the Wiener filter (matched filter) analysis. To correctly propagate the uncertainty in the covariance matrix estimation arising from the finite number of mocks used, we use the prescription given by \cite{Sellentin:2016} and calculate the final likelihood $P(A_L)$ as
\begin{equation}
    \label{eq_SH_likelihood}
    P(A_L) = \frac{\bar{c}_p \vert\mathbf{C}\vert^{1/2}}{\left[1 + \frac{\chi^2}{N_m-1}\right]^{N_m/2}},
\end{equation}
where 
\begin{equation}
    \label{eq_cp}
    \bar{c}_p = \frac{\Gamma\left(\frac{N_m}{2}\right)}{\left[\pi(N_m-1)\right]^{p/2}\Gamma\left(\frac{N_m-p}{2}\right)}.
\end{equation}

\subsection{Modeling galaxy bias within voids}
\label{sec_bias}

Measurement of the void lensing signal gives us information on the underlying matter distribution within these regions. It is interesting to compare this to the convergence profile that would be predicted from direct observation of the distribution of visible galaxies, combined with a naive assumption of a constant linear galaxy bias within voids. Denoting the mean galaxy overdensity profile in void regions as $\delta_g(r)$, this assumption allows us to relate it to the mean matter density profile by $\delta_g(r)=b\delta(r)$, where $b$ is the galaxy bias. Generalizing Eq. \ref{eq:kappa model} then gives
\begin{multline}
    \label{eq:kappa lin. bias}
    \kappa^\mathrm{bias}(\theta)=\frac{3\Omega_{m}H_0^2}{2c^2b}\int \frac{\mathrm{d}n_v(z)}{\mathrm{d}z}\\
     \int \frac{\chi\left(\chi_s-\chi\right)}{\chi_s}\frac{\delta_g(\theta,\chi)}{a}\mathrm{d}\chi\mathrm{d}z\,,
\end{multline}
where we have introduced an additional integral over the redshift distribution of the void lenses.

The galaxy overdensity $\delta_g(r)$ is mathematically identical to the monopole of the void-galaxy cross-correlation function, sometimes also denoted as $\xi^{vg}_0(r)$. We measure this monopole as a function of the void-galaxy separation $r$ for each void bin in $\lambda_v$ using a modified version of the \texttt{CUTE} correlation function code \citep{Alonso:CUTE}\footnote{The modified code is available from \url{https://github.com/seshnadathur/pyCUTE}.} through an implementation of the \citet{Landy:1993} estimator,
\begin{eqnarray}
    \label{eq:LS estimator}
    \delta_g(r) & \equiv & \xi^{vg}_0(r) \\ \nonumber
    &=& \frac{D_v D_g(r) - D_v R_g(r) - D_g R_v(r) + R_v R_g(r)}{R_v R_g(r)}\,.
\end{eqnarray}
Here each term $XY(r)$ represents the number of pairs of objects between populations $X$ and $Y$ within the given separation bin, normalized by the total number of pairs, $N_XN_Y$, where $N_X$ is the number of objects in population $X$. The populations $D_v$ and $D_g$ are the BOSS void and galaxy samples, and $R_v$ and $R_g$ are corresponding random (unclustered) catalogs of points which match the survey selection function, geometry and systematic effects present in the data, and contain 50 times as many points as the data. The ``galaxy'' random catalog $R_g$ is provided with the BOSS public data release. Further details of the construction of the ``void'' random catalog $R_v$ and the measurement of the cross-correlation are given in \cite{nadathur19}.

To convert measurements of $\delta_g(r)$ to model convergence profiles via Eq. \ref{eq:kappa lin. bias}, we assume a value for the bias $b=b_\mathrm{CMASS}=2$. This is in line with the mean value of the linear bias deduced from CMASS galaxy clustering measurements \citep{gilmarin15} and used in BOSS analyses \cite{Alam:2017}. This bias model is thus exactly analogous to the model used by \citet{alonso18} to predict the tSZ signal from voids.

There are many reasons to expect that this na\"ive assumption might not hold, which include possible environmental dependence of the bias and a statistical selection bias \citep{Nadathur:2019a} that we discuss further in \S \ref{sec_bias_fitting}. Our purpose here is to examine whether the failure of this assumption can be directly seen in the data. To this end, we repeat the fitting procedure described in \S\ref{sec_fitting} using the models of Eq. \ref{eq:kappa lin. bias} to determine the relative lensing amplitude, denoted $A_L^\mathrm{naive}$ in this case to distinguish it from that obtained from the simulation template models. The procedure could equivalently be viewed as fitting for the appropriate value of the inverse bias $1/b$.

\section{Results}
\label{sec_results}

\begin{figure*}
\centering
\includegraphics[width=\textwidth, keepaspectratio]{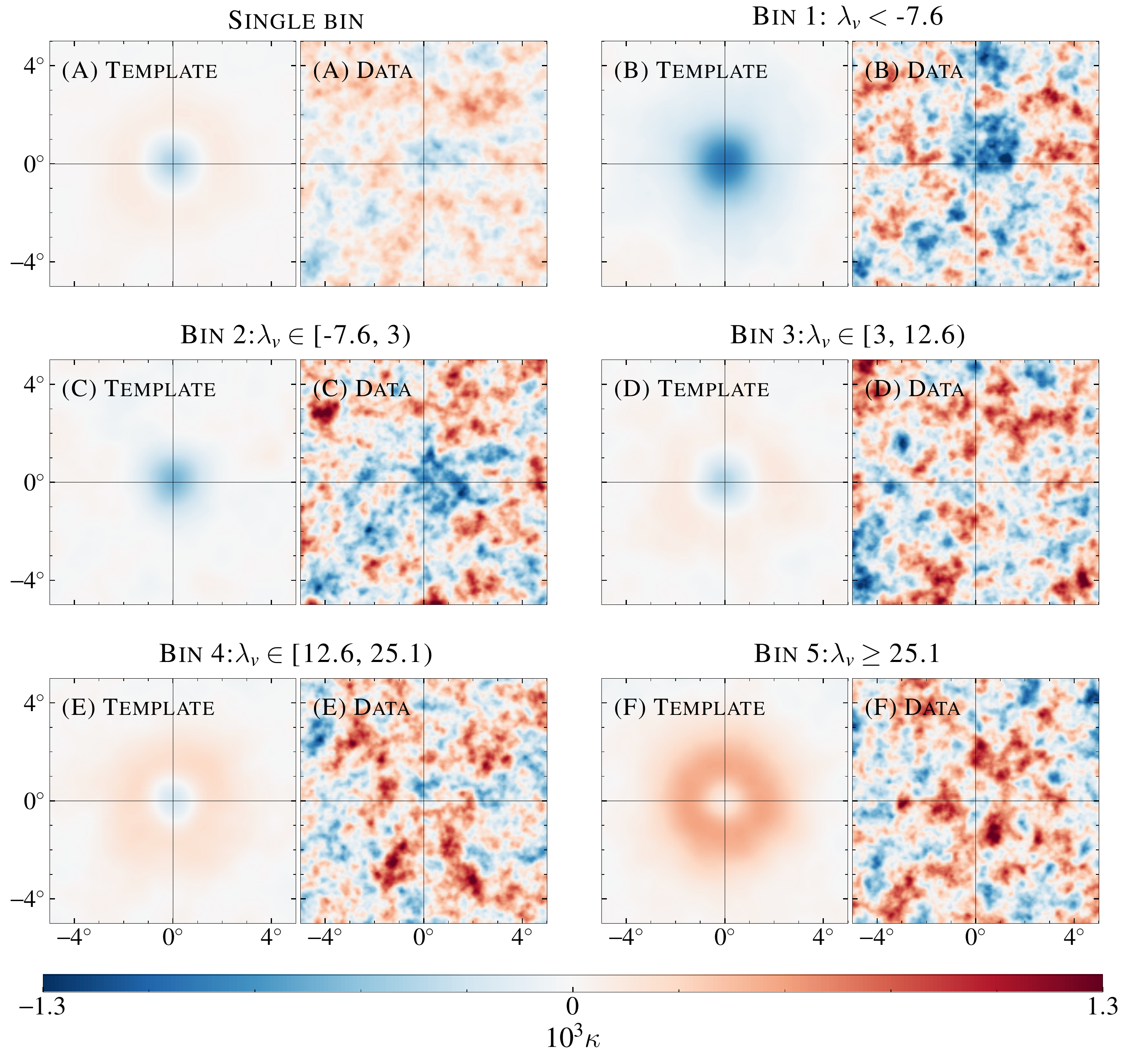}
\caption{
Mean lensing convergence ($10^{3} \kappa$) signal at the locations of BOSS voids, obtained by stacking patches extracted from the Wiener-filtered \planck{} lensing map. The Wiener-filtered templates (left) and data (right) are shown next to each other to highlight the resemblance between the structures in them. 
Panel (A) shows the result for the stack of all \howmanyvoidsintotal{} voids taken together. Panels (B) through (F) show the stacked results for different sub-populations of voids in bins of increasing $\lambda_v$ as indicated.   
Upon comparing panel (A) with others, it is evident that a single bin stack of all voids is sub-optimal for the void lensing detection. 
The positive $\kappa$ signal in panel (F) is due to the projection effect mentioned in \S\ref{sec_calibration}, combined with the effect of the Wiener filter. 
}
\label{fig_wiener_filter_stacks}
\end{figure*}

\begin{figure*}
\centering
\includegraphics[height=0.96\textwidth, keepaspectratio]{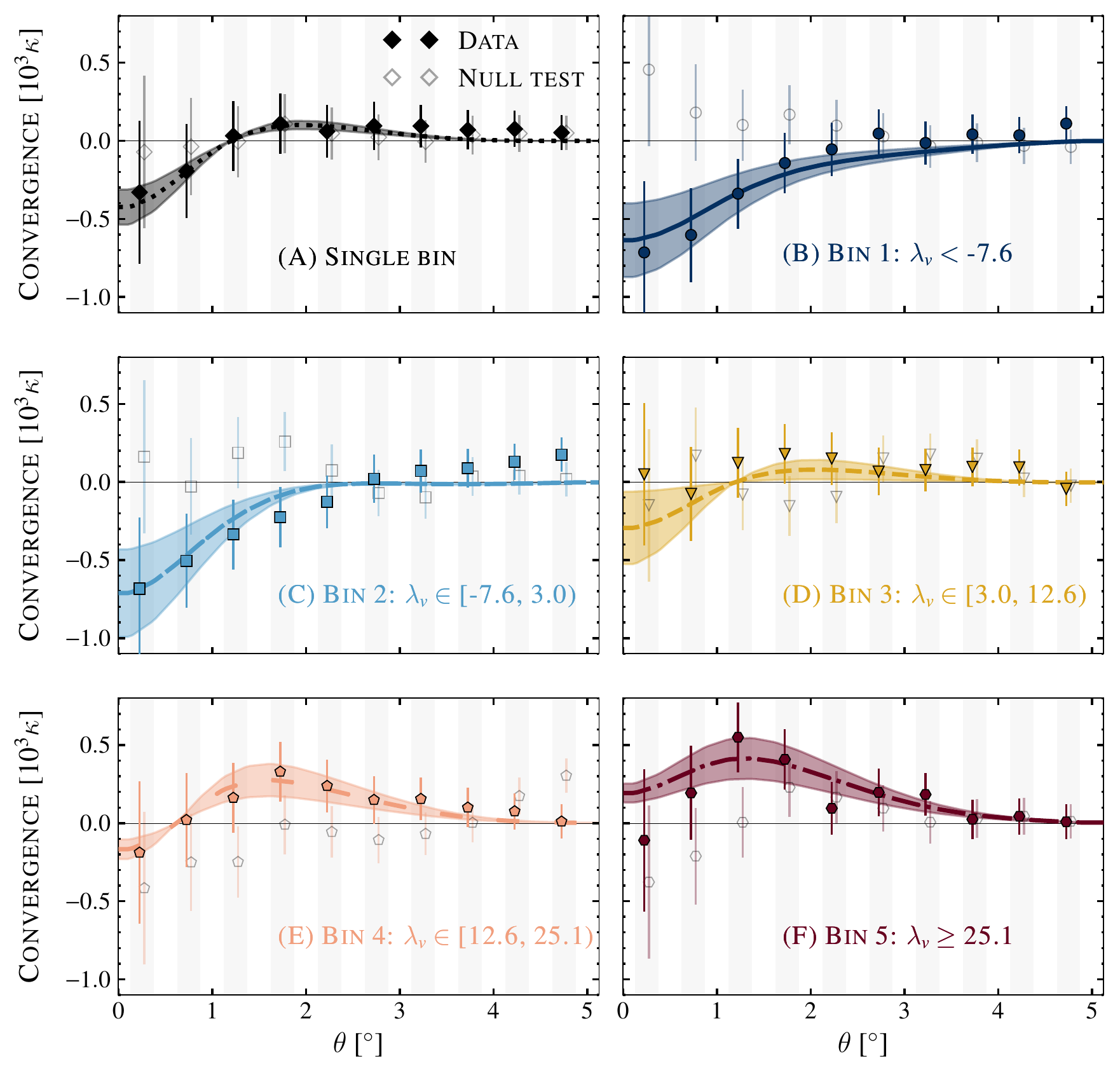}
\caption{
Stacked radial profiles $\hat\kappa^\mathrm{WF}(\theta)$ of the Wiener-filtered CMB lensing convergence around BOSS void center locations are shown as filled data points, with error bars derived from the diagonal terms of the covariance matrix. The stacked result for all voids in a single bin is shown in panel (A), with panels (B) through (F) showing stacks in bins of increasing void $\lambda_v$. The open data points show null test results obtained from stacks centered at the locations of voids in mock catalog. 
The solid curves show the lensing templates from simulation, after convolution with the Wiener filter and scaled by the best-fit lensing amplitude $\Al$ for each individual bin. The shaded regions indicate the $68\%$ C.L. posterior range for $\Al$ in each case. 
To avoid visual clutter, data has been re-binned to $\Delta \theta = 0.5^{\circ}$ from the bin size of $\Delta \theta = \thetabinsize$ used in fitting. The positive $\kappa$ signal in panel (F) is due to the projection effect mentioned in \S\ref{sec_calibration}, combined with the effect of the Wiener filter. 
}
\label{fig_lambda_5bins_wiener_filter_results_radial_profile}
\end{figure*}

\begin{figure}
\centering
\includegraphics[width=0.47\textwidth, keepaspectratio]{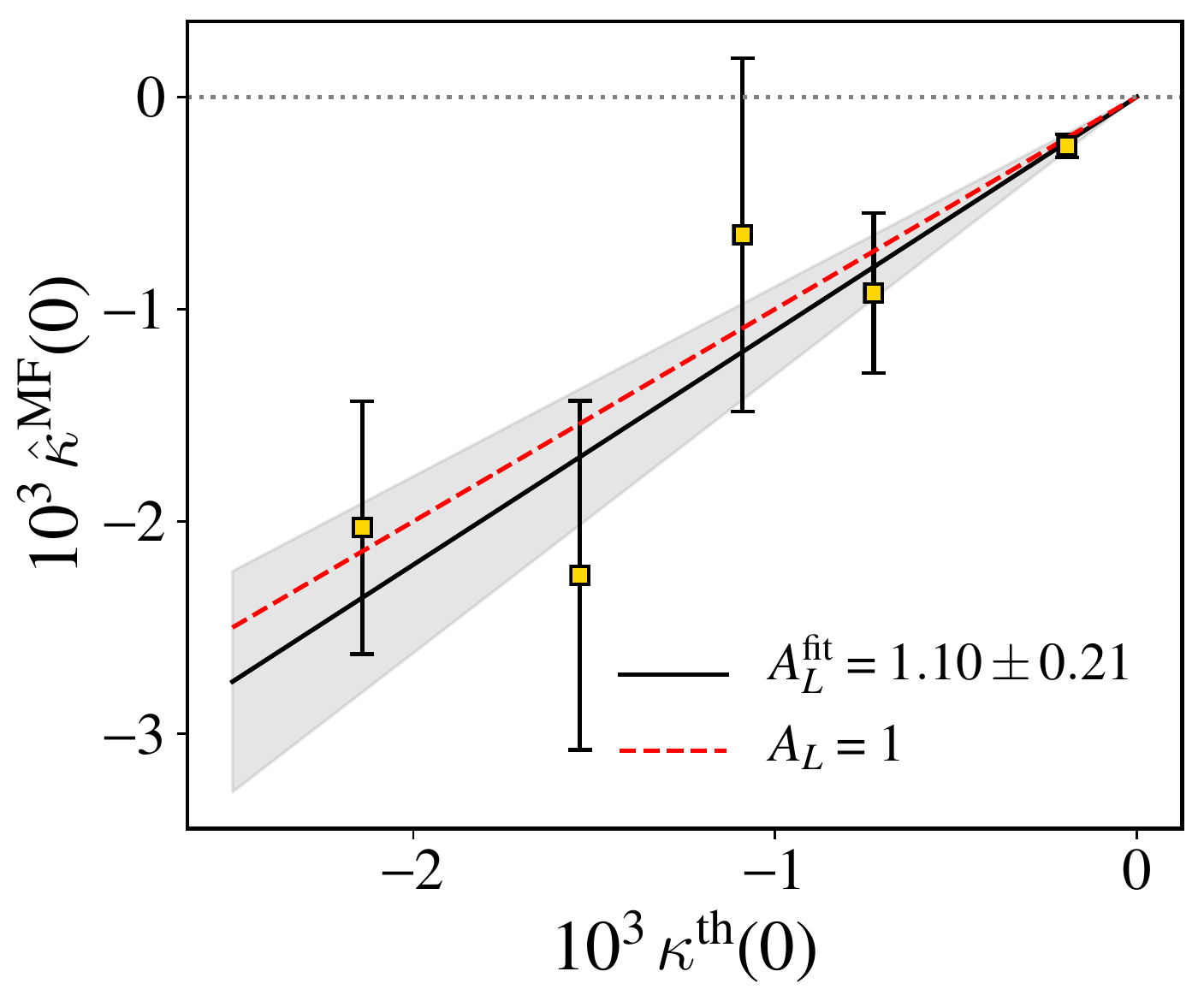}
\caption{The stacked void lensing contributions $10^3\hat\kappa^\mathrm{MF}(0)$ measured in the matched-filtered \planck{} reconstructed lensing convergence maps in each void $\lambda_v$ bin, compared to the model values obtained from calibration with lensing simulations as described in the text. Error bars show the square roots of the diagonal entries in the full covariance matrix. Bin values of $\lambda_v$ increase from left to right. The solid black line shows the best-fit value for the lensing amplitude $A_L$ derived from this data, and the shaded region shows the $68\%$ C.L. posterior range on $A_L$. The red dashed line shows the fiducial value $A_L=1$.
}
\label{fig_mf_al_fit}
\end{figure}

For illustrative purposes, Figure~\ref{fig_wiener_filter_stacks} shows the stacked patches extracted from the Wiener-filtered \planck{} $\kappa$ map at void locations. 
As in the case of Figure \ref{fig_lambda_5bins_stacks_takahashi_sims}, we show the stack for all voids taken together (panel A) and for the individual $\lv$ bins (panels B through F). 
We show both the Weiner-filtered templates (left) and the data (right) next to each other to highlight the close resemblance between the structures in the two panels. 
Even by eye, the $\kappa<0$ central de-magnification region in panel (B) and the $\kappa>0$ ring in panel (F) are clear. 
Our quantitative analysis is however performed not directly on these stacks but on the azimuthally-averaged profiles $\kappa(\theta)$ extracted from them. These are shown in Figure~\ref{fig_lambda_5bins_wiener_filter_results_radial_profile}, together with the best-fit template profiles in each case. For visual clarity, the profiles are shown rebinned into bins of width $\Delta\theta=0.5^\circ$, though fits are performed with the original binning. Error bars in this figure are derived from the diagonal elements of the covariance matrix, but due to significant off-diagonal contributions neighboring bins are correlated with each other. For the matched-filter analysis, we plot the observed stack values $\hat\kappa^\mathrm{MF}$ against the model expectations from the templates in Figure \ref{fig_mf_al_fit}.

\begin{figure}
\centering
\includegraphics[width=0.47\textwidth, keepaspectratio]{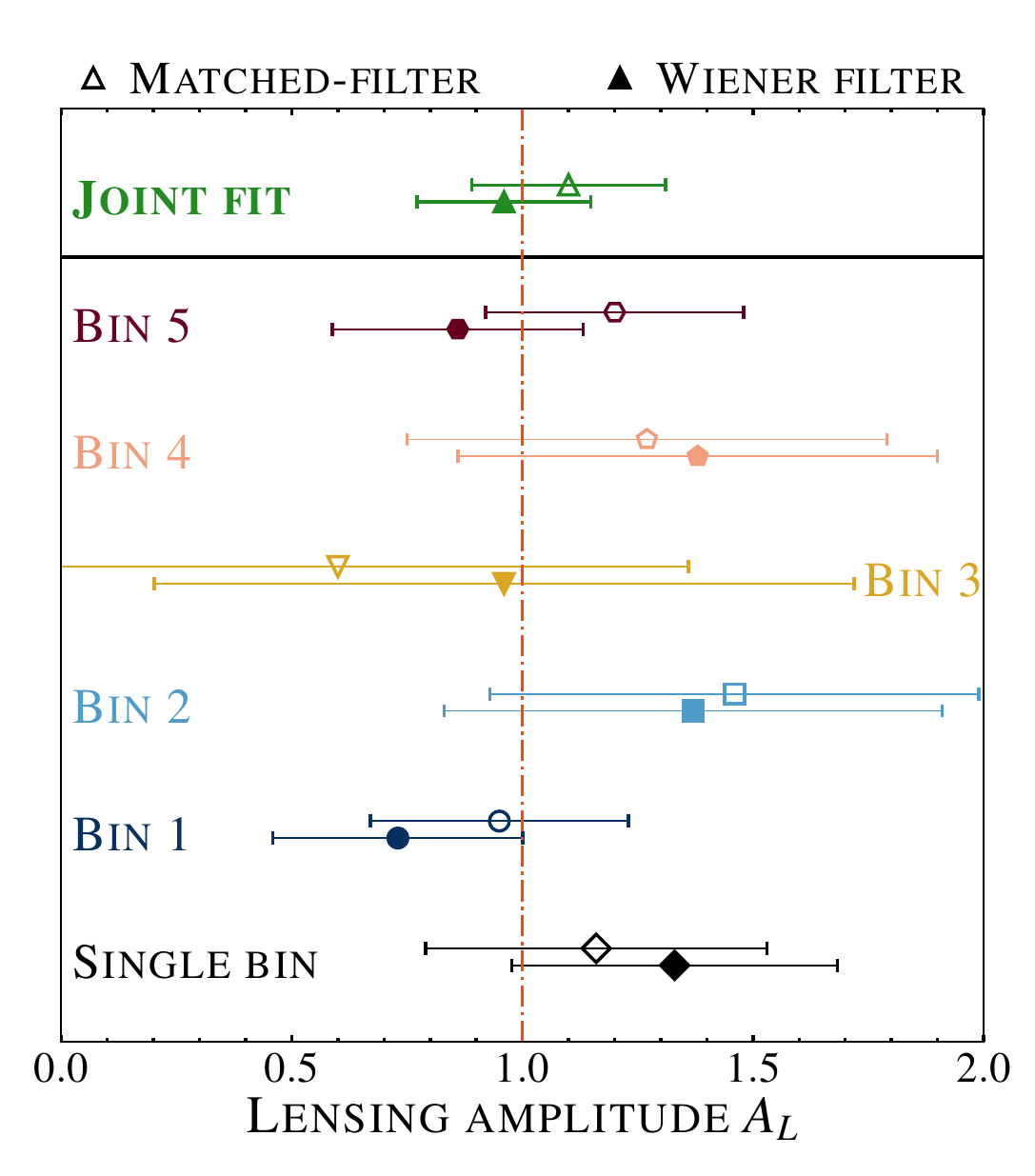}
\caption{The best-fit amplitudes $\Al$ for the measured void lensing signals relative to the templates from the \takasims{} ($\Al=1$, shown by the red dash-dotted vertical line). Open and filled points denote the matched-filter and Wiener-filter analyses, respectively. Our headline results, labelled ``{\bf joint fit}'', are obtained from jointly fitting for $A_L$ to the measured lensing signal in all five $\lambda_v$ bins and are shown in the top panel as green triangles.
For reference, the results from individual bins and the single-bin stack of all voids together are shown in the bottom panel, and are in good agreement with the results in the top panel.}
\label{fig_al_fit}
\end{figure}

The results obtained for fits to the void lensing amplitude $A_L$ using the two filtering and stacking approaches described in \S\ref{sec_methods} are summarised in Figure~\ref{fig_al_fit}. Our headline results, obtained for the joint fit to voids in all $\lambda_v$ bins, are $\Al=0.97\pm0.19$ using the Wiener filtering, and $\Al=1.10\pm0.21$ for the matched filters. These results are in excellent agreement with each other and with the expectation $\Al=1$ for the lensing templates derived from simulation. They represent rejection of the no-lensing hypothesis at the $\wfcombinedsnr \sigma$ and $\mfcombinedsnr \sigma$ significance levels respectively. 
The consistency between the two filtering approaches demonstrates that the assumption of template profiles used in designing the matched filters does not introduce any significant biases. 

In addition to these headline results, we also fit for $\Al$ separately in each individual void bin, and for the stack of all voids together in a single bin. 
These fits to the Wiener-filtered stacks in individual bins are shown as the model curves in Figure~\ref{fig_lambda_5bins_wiener_filter_results_radial_profile}. 
To quantify the goodness of fit in each bin we compute the probability-to-exceed (PTE) values, which are 55\%, 69\%, 45\%, 76\%, 20\% for the individual bins; 26\% for the single bin all-void stack; and 44\% for the combined fit. 
The $\Al$ fit results are also summarised in Table~\ref{tab_results} and Figure~\ref{fig_al_fit}. In each case the results obtained are consistent with the combined fit to within the increased statistical errors, showing that there are no significant outliers among the void sub-populations. 

\begin{deluxetable*}{c|c|ccccccc}
\tablecaption{Amplitude of void CMB-lensing signal detected in this work with respect to the templates from the \takasims{}, for different analysis choices.}
\label{tab_results}
\tablehead{
\multirow{2}{*}{Type} & \multirow{2}{*}{Filter} & \multicolumn{7}{c}{Lensing amplitude $\Al$} \\
\cline{3-9}
& & \colhead{Single bin} & \colhead{Bin 1} & \colhead{Bin 2} & \colhead{Bin 3} & \colhead{Bin 4} & \colhead{Bin 5} & \colhead{{\bf Joint fit}}}
\startdata
\hline
\multirow{2}{*}{Baseline} &  Wiener & $\wfallvoidsAl$ & $\wffirstbinAl$ & $\wfsecondbinAl$ & $\wfthirdbinAl$ & $\wffourthbinAl$ & $\wffifthbinAl$ & $\wfcombinedAl$ \\
& Matched & $\mfallvoidsAl$ & $\mffirstbinAl$ & $\mfsecondbinAl$ & $\mfthirdbinAl$ & $\mffourthbinAl$ & $\mffifthbinAl$ & $\mfcombinedAl$ \\\hline\hline
\multirow{2}{*}{tSZ-nulled} & Wiener & $\noSZwfallvoidsAl$ & $\noSZwffirstbinAl$ & $\noSZwfsecondbinAl$ & $\noSZwfthirdbinAl$ & $\noSZwffourthbinAl$ & $\noSZwffifthbinAl$ & $\noSZwfcombinedAl$ \\
& Matched & $\noSZmfallvoidsAl$ & $\noSZmffirstbinAl$ & $\noSZmfsecondbinAl$ & $\noSZmfthirdbinAl$ & $\noSZmffourthbinAl$ & $\noSZmffifthbinAl$ & $\noSZmfcombinedAl$ \\\hline\hline
Null test & Wiener & $\nullwfallvoidsAl$ & $\nullwffirstbinAl$ & $\nullwfsecondbinAl$ & $\nullwfthirdbinAl$ & $\nullwffourthbinAl$ & $\nullwffifthbinAl$ & $\nullwfcombinedAl$ \\
\enddata
\tablecomments{Our headline results, labelled ``{\bf joint fit}'', are obtained from jointly fitting for $A_L$ to the measured lensing signal in all five $\lambda_v$ bins.
}
\end{deluxetable*}

\subsection{Comparison to previous work}
\label{sec_comparison}
Two previous works \citep{cai17, vielzeuf19} have studied the lensing imprints of voids on the CMB, although both report lower significance detections, at just over the $3\sigma$ level. \citet{vielzeuf19} used a void catalog constructed from the DES Year 1 data. 
Since these data have only photometric redshifts, the redshift smearing effect favours finding voids in the projected 2D density field rather than the 3D field used here. 
This leads to preferentially selecting voids that are elongated along the line of sight, and so each void has an enhanced lensing effect compared to those studied here \citep[in this context, see also, e.g.,][]{davies18}.
However the final void catalog then has a factor of $\sim6\times$ fewer objects than ours. 

In contrast, \citet{cai17} use a void catalog that is very similar to ours, and is also derived from the BOSS DR12 CMASS sample. Although we do use a later update of the lensed-CMB data, viz. the \planck{} 2018 release rather than 2015, the primary reason for the improved statistical significance reported in this work is our use of the novel stacking strategy recommended by \citet{nadathur17}, binning voids by the combination of their density and size encapsulated in parameter $\lambda_v$. As predicted by \citet{nadathur17}, we find that the two extreme $\lambda_v$ bins produce the best detection significance, but that their lensing imprints would partially cancel each other if included in the same stack. 
The strategy followed by \citet{cai17} of instead rescaling lensed-CMB patches in the stack in proportion to the void apparent angular scale is less efficient. 
Indeed comparison of the \citet{cai17} result with the ``single bin'' results reported here shows that rescaling on the basis of void size produces only marginally better results than simply stacking all patches together irrespective of void properties. 
We explicitly checked this by testing an alternative method of rescaling the patches based on the void sizes before stacking without reference to $\lambda_v$, as done by \citet{cai17}, and found a resulting detection significance of $\sim3.5\sigma$ that closely matches the previous results by those authors and the single-bin results quoted in Table 2.

\subsection{Tests for systematics}
\label{sec_validation} 

We performed a random-positions null test for systematics in our measurement by replacing the BOSS voids used in the data measurement by voids drawn from a randomly selected mock void catalog from the \takasims. This preserves the effects of the clustering of void positions and their overlap, and ensures that patches are drawn from within the same BOSS footprint in the sky (which may be important in case of inhomogeneous noise properties or large-scale modes affecting the \planck{} $\kappa$ map). 
However, the locations in the mock catalog do not have any correlation with the \planck{} CMB lensing and so should return a null signal.
The results obtained are shown as open data points in Figure~\ref{fig_lambda_5bins_wiener_filter_results_radial_profile} and in Table~\ref{tab_results}, and are consistent with no lensing signal, as expected.

We then repeated all the measurement and fitting procedures described above for two other cases. 
First, we checked the robustness of the results from Wiener and matched-filter analysis by eliminating both the filters; second, we performed a more conservative analysis to check the effect of removing large-scale modes $L<8$ from the \planck{} $\kappa$ map.
The results obtained for these cases are
\begin{eqnarray}
    \label{eq_hpf_Al_results}
    A_L &=& \nfcombinedAl\;\;(L \le 2048,\;\mathrm{no\,filter}) \\ \nonumber
    &=& \wfcombinedlargescalesfilteredAl\;\;(8 \le L \le 2048,\; \mathrm{Wiener\,filter)} \\ \nonumber
    &=& \mfcombinedlargescalesfilteredAl\;\;(8 \le L \le 2048,\; \text{matched-filter)}.
\end{eqnarray}
These numbers are entirely consistent with the headline results obtained above, indicating no significant contamination from the inclusion of the large-scale modes or the filters used. We also note that, for the current noise levels, the constraints on $\Al$ obtained from Wiener- or matched-filters are not significantly better than the result without any filtering.  
However, the optimal filters used here will be important for measurements with the next generation low-noise CMB datasets. 

As mentioned in \S\ref{sec_datasets}, an important potential contaminant of the void lensing signal comes from tSZ signals. \citet{alonso18} already reported a $3.4\sigma$ detection cross-correlation of BOSS voids with tSZ signal from a stacking analysis using \planck{} Compton-$y$ maps \citep{hurier13, planck16_tsz}. tSZ signal is a known contaminant of the lensing convergence reconstruction from CMB temperature data \citep{vanengelen13, madhavacheril18}. We therefore repeated our entire analysis pipeline on the \planck{} convergence map reconstructed from tSZ-deprojected temperature data only. This map produced from tSZ-nulled CMB map is noisier than the fiducial lensing map produced using a MV combination of data from multiple channels and as a result the detection sensitivity expected is lower than the fiducial case. The resulting $A_L$ constraints are included in Table \ref{tab_results}, with final combined fit values
\begin{eqnarray}
    \label{eq_hpf_Al_results}
    A_L &=& \noSZwfcombinedAl\;\;\mathrm{(Wiener\,filter)} \\ \nonumber
    &=& \noSZmfcombinedAl\;\;\text{(matched-filter)}.
\end{eqnarray}
These have a higher uncertainty as expected, but are entirely consistent both with our headline results and with $A_L=1$, indicating no significant tSZ contamination.

We do not directly test for contamination from the kinematic SZ signal arising due to the motion of galaxies \citep{ferraro18}, but given the absence of tSZ contamination this effect is also expected to be negligible. Finally, we also neglect the possibility of contamination from cosmic infrared background (CIB) emission \citep{vanengelen13, osborne14, schaan18}, because the CIB is sourced primarily by high-redshift galaxies at $z\ge2$ which should have only a small correlation with the BOSS galaxies at redshifts $z\lesssim0.7$.

\subsection{Testing the linear bias model in voids}
\label{sec_bias_fitting}

The results above demonstrate that the void CMB lensing signal is seen at high significance and is completely consistent with templates calibrated from the lensing simulations of \citet{takahashi17}. We now replace these simulation templates with those constructed in \S\ref{sec_bias} by naively assuming a constant linear galaxy bias $b=b_\mathrm{CMASS}=2$ \citep{gilmarin15, Alam:2017} is valid within voids. With the exception of the change of templates, the fitting procedure remains the same, although for simplicity in this case we only perform the Wiener-filtered stacking. To avoid confusion, we denote the lensing amplitude measured in this case as $A_L^\mathrm{naive}$. The result obtained from the joint fit to all bins is
\begin{equation}
    \label{eq_Al_naive_bias_fit}
    A_L^\mathrm{naive} = \wfcombinednaiveAl\,.
\end{equation}
Once again the results from fits in individual $\lambda_v$ bins are entirely consistent with the joint result with larger uncertainties, but are not shown explicitly for simplicity.

This value is discrepant with the expectation $A_L=1$ at $2.8\sigma$, indicating that the observed lensing effect of voids is significantly smaller than would be expected from the assumption of constant linear galaxy bias within voids matching that the overall CMASS sample. This result can be rephrased in terms of a constraint on the effective bias $b_\mathrm{eff}=\delta_g/\delta$ within voids: if this constant proportionality held for all $r$ and for all voids, the implied value of the bias would be $b_\mathrm{eff}^{-1}=\wfallinvvoidsbias$, equivalent to a mean galaxy bias of $b_\mathrm{eff}=\wfallvoidsbiasnoerrorbar$ and inconsistent with the results from galaxy clustering at the same $\sim3\sigma$ level.

These results are consistent with those of \citet{Nadathur:2019a}, who show that in simulations, the galaxy distribution and matter distribution in voids are not in general related by a simple linear bias relationship, and that assuming such a relationship leads to an overestimate of the matter deficit within voids. Our results are also consistent with previous work on the tSZ emission profiles of BOSS voids by \citet{alonso18}. These authors built a theory model for the tSZ signal assuming linear bias, in a manner entirely analogous to our method in \S\ref{sec_bias}. On fitting the model to data, they find a relative amplitude factor $\alpha_v=0.67\pm0.2$, close to $2\sigma$ smaller than the expectation $\alpha_v=1$, and entirely consistent with our result for $A_L^\mathrm{naive}$ given their larger uncertainties. Using lensing shear measurements for a different catalog of voids obtained from galaxy samples in DES \mbox{Year 1} with photometric redshifts only, \citet{Fang:2019} report better agreement with the assumption of a linear relationship between the void matter and galaxy density profiles, but again with an effective bias factor that is in excess of that determined from galaxy clustering for the vast majority ($\sim85\%$) of voids (e.g., see their Figure 14).

Contributions to this apparent lack of matter deficit within voids could come from several sources, including the possibility that the relationship between mass and luminous galaxies is fundamentally different in low-density environments. However, as pointed out by \citet{Nadathur:2019a}, the single biggest contribution is likely to be from a simple statistical effect. Irrespective of the true mean relationship between mass and galaxies in low-density environments, there will necessarily be significant scatter around this mean due to shot noise fluctuations in the galaxy distribution. The fact that voids are selected on the basis of searching for regions of low galaxy density then \emph{necessarily} introduces a statistical bias in the observed mass-to-light ratio in these selected regions, which works in the direction of these voids containing smaller matter deficits than predicted from their galaxy content. In simulation this effect leads to deviations of up to $\sim25\%$ from the expected $\delta(r)$ \citep{Nadathur:2019a}. The only way to completely eliminate this selection bias is to define voids not on the basis of the observed galaxy density field, but directly from the matter density $\delta$. This might be possible by adapting void-finding algorithms to operate on lensing convergence maps rather than the galaxy field, but we leave this to future work. 

\section{Void-CMB lensing measurements with future data}
\label{sec_future_prospects}

In this work we used only lensing reconstruction results from \planck. However, future CMB data from Advanced ACT (AdvACT, \citealt{henderson16}), CMB Stage IV (CMB-S4, \citealt{S4_DSR_19}), and Simons Observatory (SO, \citealt{SO19}) are expected to lower the lensing reconstruction noise by an order of magnitude. The noise in the polarization channels of these experiments will also be low enough to allow efficient lensing reconstruction from CMB polarization-only data, eliminating lensing systematics induced by SZ contamination \citep{hall14, yasini16} and emissions from extragalactic foregrounds that are largely unpolarized \citep{datta18, gupta19}, that affect temperature-based CMB lensing maps. These surveys will not be full-sky, but will scan roughly 40\% of the southern sky, giving sufficient overlap for excellent synergies with optical surveys including DES \citep{desoverview_16}, the Dark Energy Spectroscopic Instrument (DESI, \citealt{DESI16_science_paper_I}), Euclid \citep{euclid10}, and the Large Synoptic Survey Telescope (LSST, \citealt{lsst09})

\begin{deluxetable}{l|c|c|ccc}
\tablecaption{Expected \snr{} of BOSS voids using different CMB lensing estimators for the future CMB surveys.}
\label{tab_future_prospects}
\tablehead{
\multirow{2}{*}{Experiment} & $\Delta_{T}$ & \multirow{2}{*}{$L_{\rm max}$} & \multicolumn{3}{c}{Void lensing \snr{}}\\
\cline{4-6}
& [$\mu {\rm K}^{\prime}$] & & \colhead{MV} & \colhead{TT} & \colhead{MVpol}
}
\startdata
\hline
Third generation & 7.0 & 3000 & 9.3 & 7.5 & 8.2 \\\hline
CMB-S4 & 2.0 & 4000 & 10.7 & 9.5 & 10.1  \\\hline
SO-baseline & 10.0 & 4000 & 9.4 & 8.4 & 8.0 \\\hline
SO-goal & 6.3 & 4000 & 9.7 & 9.0 & 9.1 \\\hline
\enddata
\end{deluxetable}

In this section, we forecast the detection sensitivity achievable for the void-CMB lensing measurement for a current ``third generation'' and future (CMB-S4 and SO) experiments. To allow an easy comparison to the results from \planck{}, in making these forecasts we assume that the void population used for such a detection has the same properties as the BOSS voids used in the current work. That is, we assume the number of void lenses available for the measurement is fixed, and that their matter profiles and redshift distribution are close enough to those of BOSS voids that their predicted lensing contributions will be similar. 
We assume a noise level of $\Delta_{T} = 7.0\ \mu \mathrm{K}^{\prime}$ ($\Delta_{P} = \sqrt{2}\Delta_{T}$) for the third generation experiment 
and $\Delta_{T} = 2.0,\,10.0$ and $6.3\ \mu \mathrm{K}^{\prime}$ for CMB-S4, SO-baseline, and SO-goal respectively. 
A common beam of $1.^{\prime}5$ at 150 GHz was assumed for all experiments which corresponds to a telescope with a primary dish size of 6m. The expected lensing noise curves were computed for the above experiments assuming a maximum lensing multipole $L_{\rm max} = 3000$ for the third generation experiment and $L_{\rm max} = 4000$ for other experiments. A higher $L_{\rm max}$ was assumed for the next generation experiments as they are being designed to have a broader frequency coverage compared to current experiments for an efficient suppression of extragalactic foreground signals.
We obtained the lensing curves for the temperature-only (TT), MV combination of all the five temperature and polarization-based lensing estimators (TT, TE, EE, EB and TB), and MV combination of the two polarization-only (MVpol) estimators. 

The forecast achievable signal-to-noise for the void lensing measurements in these data are summarised in Table~\ref{tab_future_prospects}. All the future experiments can reach detection sensitivities far in excess of those achieved with \planck{}, exceeding $10\sigma$ for CMB-S4. The lensing \snr{} from polarization-only channels is equal to or better than temperature-based estimation for experiments with $\Delta_{T} \le 7.0\ \mu \mathrm{K}^{\prime}$. These forecasts can be regarded as conservative because, given the capabilities of surveys such as DESI, Euclid and LSST, the number of voids used for future measurements is expected to increase.

\section{Conclusion}
\label{sec_conclusion}

We have reported a high-significance detection, at the $\mfcombinedsnr \sigma$ level, of the gravitational lensing effect in \planck{} data of cosmic voids found in the BOSS CMASS galaxy sample. The measured signal amplitude and shape is consistent with the lensing templates we derived from mock void catalogues and full-sky CMB lensing maps in a suite of 108 simulations created by \citet{takahashi17}. We tested our measurement pipeline with two theoretically-motivated filtering strategies to reduce the noise in the \planck{} $\kappa$ map, using optimal matched filters and a Wiener filter, and obtained consistent results with both. The matched filter technique is designed to maximise the lensing \snr{}, but the filter design assumes detailed knowledge of the void lensing template, which the Wiener filter approach does not. The consistency of the two methods therefore serves as a good cross-check that the shape of the lensing signal in the data indeed matches the simulation well. The \snr{} we report for the void CMB lensing detection here is significantly higher than that reported in two previous studies \citep{cai17, vielzeuf19}. This difference is primarily due to improvements in the stacking methodology introduced here, as well as to improvements in the datasets and filtering.

Our improved sensitivity allows us to probe the relationship between the matter, $\delta(r)$, and galaxy, $\delta_g(r)$, density profiles around void locations.  Using direct measurement of void $\delta_g(r)$ from the galaxy data, we tested the hypothesis that this relationship is linear, with the same constant bias as determined from galaxy clustering analyses of the CMASS sample, and rule it out at $\sim 3\sigma$ significance. 
This hypothesis overpredicts the amplitude of the void lensing effect by close to 40\%; equivalently, if the galaxy bias relationship within voids is truly linear, this bias must be $\sim60\%$ larger than the value $b_\mathrm{CMASS}=2$ deduced from galaxy clustering. This result is not unexpected due to a strong statistical selection bias arising from the void identification that has been confirmed in simulations \citep{Nadathur:2019a}, potentially in combination with the impact of additional non-linear biasing. It is also consistent with the result obtained (albeit with lower significance) in the context of the void-tSZ cross-correlation by \citet{alonso18}, who also used voids in the CMASS sample. While the linear bias model predictions fail, the predicted lensing profiles obtained from directly tracing the total matter content of voids in simulation is in very good agreement with observation.

Finally, we forecast the signal-to-noise achievable for void-CMB lensing measurements with future data from current and next generation experiments like CMB-S4 and SO. These data will allow the detection of the void lensing signal at significance far exceeding that achieved with \planck{}, and with negligible systematics. The precision obtained from these measurements of $\kappa(\theta)$ will then enable inversion to determine the matter profiles $\delta(r)$ directly from data. This method could then replace the current necessity of calibrating these profiles against simulation results for use in other measurement of void dynamics \citep[e.g.][]{nadathur19}. The direct determination of $\delta(r)$ will also be an important factor in the use of voids to for cosmological applications, such as probing the sum of neutrino masses $\Sigma m_\nu$ and testing modified gravity models.

\acknowledgments
We thank the anonymous referee for useful suggestions and comments.
We thank Ryuichi Takahashi for correspondence regarding the  \cite{takahashi17} simulations. 
We acknowledge the Centro de Ciencias de Benasque Pedro Pasqual and the organisers of the workshop ``Understanding Cosmological Observations'' in July 2019, where this work was started. 
SR and NW acknowledge support from NSF grants AST-1716965 and CSSI-1835865. 
SN is supported by UK Space Agency grant ST/N00180X/1. BDS acknowledges support from an STFC Ernest Rutherford Fellowship, an Isaac Newton Trust Early Career Grant, and from the European Research Council (ERC) under the European Union's Horizon 2020 research and innovation programme (Grant agreement No. 851274). This work has made use of the \texttt{Hoffmann2} cluster at UCLA and the UK \textsc{Sciama} High Performance Computing cluster supported by the ICG, SEPNet and the University of Portsmouth.

This work has made use of public data from the SDSS-III collaboration. Funding for SDSS-III has been provided by the Alfred P. Sloan Foundation, the Participating Institutions, the National Science Foundation, and the U.S. Department of Energy Office of Science. The SDSS-III website is \url{http://www.sdss3.org/}. SDSS-III is managed by the Astrophysical Research Consortium for the Participating Institutions of the SDSS-III Collaboration including the University of Arizona, the Brazilian Participation Group, Brookhaven National Laboratory, Carnegie Mellon University, University of Florida, the French Participation Group, the German Participation Group, Harvard University, the Instituto de Astrofisica de Canarias, the Michigan State/Notre Dame/JINA Participation Group, Johns Hopkins University, Lawrence Berkeley National Laboratory, Max Planck Institute for Astrophysics, Max Planck Institute for Extraterrestrial Physics, New Mexico State University, New York University, Ohio State University, Pennsylvania State University, University of Portsmouth, Princeton University, the Spanish Participation Group, University of Tokyo, University of Utah, Vanderbilt University, University of Virginia, University of Washington, and Yale University.

\ifdefined\ApJsubmit
\bibliographystyle{aasjournal}
\else
\bibliographystyle{aasjournal_arxiv}
\fi
\bibliography{cmb_lensing_voids}

\end{document}